\documentclass[aps,prd,preprint,groupedaddress]{revtex4-1}
\usepackage[utf8]{inputenc}
\usepackage[english]{babel}
\usepackage[letterpaper,margin=1in]{geometry}

\usepackage[dvipsnames]{xcolor}
\usepackage{amsmath,amssymb,amsfonts,enumerate,float,graphicx,hyperref,mathrsfs,mathtools,fancyvrb,slashed,pdfpages,empheq,tabularx,caption,multirow,subcaption}

\usepackage[inline,shortlabels]{enumitem}

\allowdisplaybreaks

\usepackage{colortbl}
\usepackage{tikz,pgfplots}

\usetikzlibrary{calc,patterns,decorations.pathmorphing,decorations.markings,arrows}

\graphicspath{{./figures/}}

\usepackage[most]{tcolorbox}

\tcbset{
	frame code={}
	center title,
	left=0pt,
	right=0pt,
	top=0pt,
	bottom=0pt,
	colback=yellow!25,
	colframe=white,
	width=\dimexpr\textwidth\relax,
	enlarge left by=0mm,
	boxsep=5pt,
	arc=0pt,outer arc=0pt,
}

\newcommand\myshade{85}
\hypersetup{
	linkcolor  = violet!\myshade!black,
	citecolor  = YellowOrange!\myshade!black,
	urlcolor   = Aquamarine!\myshade!black,
	colorlinks = true,
}

\newcommand{\pp}[1]{\left ( #1 \right )}
\newcommand{\bb}[1]{\left [ #1 \right ]}
\newcommand{\cc}[1]{\left \{ #1 \right \}}
\newcommand{\vv}[1]{\left \vert #1 \right \vert}

\newcommand{\del}{\partial}

\newcommand{\bracket}[3]{\left\langle #1 \left\vert #2 \right\vert #3 \right\rangle}

\makeatletter
\newcommand\incircbin{\mathpalette\@incircbin}
\newcommand\@incircbin[2]{\mathbin{\ooalign{\hidewidth$#1#2$\hidewidth\crcr$#1\bigcirc$}}}

\makeatother

\newcommand*\dif{\mathop{}\!\mathrm{d}}

\newcommand{\prg}{\\ $  $ \\  }

\newcommand{\e}[1]{\mathrm{e}^{#1}}

\newcommand{\nn}{\nonumber\\ &}
\newcommand{\nneq}{\nonumber\\ & \hphantom{=}}
\newcommand{\nnhp}[1]{\nonumber\\ & \hphantom{#1}}
\newcommand{\ar}{\nonumber \\ & + & }
\newcommand{\ek}{\nonumber \\ & - & }
\newcommand{\carp}{\nonumber \\ & \times & }

\def\ba{\begin{eqnarray}}
\def\ea{\end{eqnarray}}
\def\bt{\begin{eqnarray*}}
	\def\et{\end{eqnarray*}}

\makeatletter
\newcommand{\pushright}[1]{\ifmeasuring@#1\else\omit\hfill$\displaystyle#1$\fi\ignorespaces}
\newcommand{\pushleft}[1]{\ifmeasuring@#1\else\omit$\displaystyle#1$\hfill\fi\ignorespaces}
\makeatother

\def\mqi{m_{q_{1}}}
\def\mqii{m_{q_{2}}}
\def\mV{m_{V}}
\def\fV{f_{V}}
\def\fVT{f_{V}^{T}}
\def\baru{\bar{u}}
\def\alphai{\alpha_{1}}
\def\alphaiii{\alpha_{3}}
\def\aipar{a_{1}^{\parallel}}
\def\aiper{a_{1}^{\perp}}
\def\aiipar{a_{2}^{\parallel}}
\def\aiiper{a_{2}^{\perp}}
\def\zetaiiipar{\zeta_{3}^{\parallel}}
\def\lambdaiiipartilde{\tilde{\lambda}_{3}^{\parallel}}
\def\omegaiiipar{\omega_{3}^{\parallel}}
\def\omegaiiipartilde{\tilde{\omega}_3^\parallel}
\def\lambdaiiipar{\lambda_{3}^{\parallel}}
\def\kappaiiipar{\kappa_{3}^{\parallel}}
\def\kappaiiiper{\kappa_{3}^{\perp}}
\def\omegaiiiper{\omega_{3}^{\perp}}
\def\lambdaiiiper{\lambda_{3}^{\perp}}
\def\zetaivpar{\zeta_{4}^{\parallel}}
\def\omegaivpartilde{\tilde{\omega}_{4}^{\parallel}}
\def\zetaivper{\zeta_{4}^{\perp}}
\def\zetaivpertilde{\tilde{\zeta}_{4}^{\perp}}
\def\kappaivpar{\kappa_{4}^{\parallel}}
\def\kappaivper{\kappa_{4}^{\perp}}
\def\thetaipar{\theta_{1}^{\parallel}}
\def\thetaiipar{\theta_{2}^{\parallel}}
\def\psiiipar{\psi_{2}^{\parallel}}
\def\thetaiper{\theta_{1}^{\perp}}
\def\thetaipertilde{\tilde{\theta}_{1}^{\perp}}
\def\thetaiiper{\theta_{2}^{\perp}}
\def\thetaiipertilde{\tilde{\theta}_{2}^{\perp}}
\def\phiiipertilde{\tilde{\phi}_{2}^{\perp}}
\def\phiiiper{\phi_{2}^{\perp}}
\def\avqi{\langle\langle Q^{(1)}\rangle\rangle}
\def\avqiii{\langle\langle Q^{(3)} \rangle\rangle}
\def\avqv{\langle\langle Q^{(5)} \rangle\rangle}
\def\psioper{\psi_{0}^{\perp}}
\def\psiopertilde{\tilde{\psi}_{0}^{\perp}}
\def\thetaoper{\theta_{0}^{\perp}}
\def\thetaopertilde{\tilde{\theta}_{0}^{\perp}}
\def\phioper{\phi_{0}^{\perp}}
\def\phiopertilde{\tilde{\phi}_{0}^{\perp}}
\def\phiiper{\phi_{1}^{\perp}}
\def\phiipertilde{\tilde{\phi}_{1}^{\perp}}
\def\psiiper{\psi_{1}^{\perp}}
\def\psiipertilde{\tilde{\psi}_{1}^{\perp}}
\def\psiiiper{\psi_{2}^{\perp}}
\def\psiiipertilde{\tilde{\psi}_{2}^{\perp}}
\def\BorelM2{M^2}

\def\phiiiparu{\phi _{2} ^{\parallel} (u)}
\def\phiiiperu{\phi _{2} ^{\perp} (u)}
\def\phiiiiparu{\phi _{3} ^{\parallel} (u)}
\def\psiiiiparu{\psi _{3} ^{\parallel} (u)}
\def\psiiiiperu{\psi _{3} ^{\perp} (u)}
\def\phiiiiperu{\phi _{3} ^{\perp} (u)}
\def\psiivparu{\psi _{4} ^{\parallel} (u)}
\def\phiivparu{\phi _{4} ^{\parallel} (u)}
\def\psiivperu{\psi _{4} ^{\perp} (u)}
\def\phiivperu{\phi _{4} ^{\perp} (u)}
\def\SS{\mathcal{S} (\alpha _{1}, \alpha _{3})}
\def\SSTilde{\tilde{\mathcal{S}} (\alpha _{1}, \alpha _{3})}
\def\VV{\mathcal{V} (\alpha _{1}, \alpha _{3})}
\def\AA{\mathcal{A} (\alpha _{1}, \alpha _{3})}
\def\TT{\mathcal{T} (\alpha _{1}, \alpha _{3})}
\def\TTi{\mathcal{T} _{1}^{(4)} (\alpha _{1}, \alpha _{3})}
\def\TTii{\mathcal{T} _{2}^{(4)} (\alpha _{1}, \alpha _{3})}
\def\TTiii{\mathcal{T} _{3}^{(4)} (\alpha _{1}, \alpha _{3})}
\def\TTiv{\mathcal{T} _{4}^{(4)} (\alpha _{1}, \alpha _{3})}

\def\Hat{\mathcal H}

\def\I{\mathcal I}

\makeatletter
\AtBeginDocument{\let\LS@rot\@undefined}
\makeatother

\begin{document}
\title{Strong vertices of doubly heavy spin-3/2--spin-1/2 baryons with light mesons in light-cone QCD sum rules}

\author{T. M. Aliev}
\email[]{taliev@metu.edu.tr}
\affiliation{Department of Physics, Middle East Technical University, Ankara 06800, Turkey}

\author{K. \c Sim\c sek}
\email[]{ksimsek@u.northwestern.edu}
\affiliation{Department of Physics \& Astronomy, Northwestern University, Evanston, IL 60208, USA}

\date{November 6, 2020}

\begin{abstract}
	In this paper, we analyze the doubly heavy spin-3/2--spin-1/2 baryons and light meson vertices within the method of light-cone QCD sum rules. These vertices are parametrized in terms of one (three) coupling constant(s) for the pseudoscalar (vector) mesons. The said coupling constants are calculated for all possible transitions. The results presented here can serve to be useful information in experimental as well as theoretical studies of the properties of doubly heavy baryons.
\end{abstract}
\maketitle 
\section{Introduction}\label{sec:1}
The quark model predicts the existence of many doubly heavy baryons. At the present time, only one of these baryons, $ \Xi _{cc}^{++} $, is discovered in experiments in the decay modes 
$ \Lambda _c^+K^-\pi^+ $ and $ pD^+K^- $ with mass $ m _{\Xi _{cc}^+} = 3518.7 \pm 17 {\rm\ MeV} $ by the SELEX Collaboration at Fermilab \cite{Mattson2002, Ocherashvili2005} and by the LHCb Collaboration \cite{Aaij2017, Aaij2018} in the decay channel $ \Xi _c^+ \pi^+ $ with the mass $ m _{\Xi _{cc}^{++}} = 3621.24 \pm 0.65 \pm 0.31 {\rm\ MeV} $. 
\prg
The main effort of experimentalists is focused on the discovery of the other members of doubly heavy baryons predicted by the quark model. From the theoretical side, these baryons provide an excellent laboratory to study their electromagnetic, weak, and strong decays for a better understanding their quark structure, check the predictions of heavy-quark symmetry as well as gain information about perturbative and nonperturbative aspects of QCD. 
\prg
The strong coupling constants of the doubly heavy baryons with light mesons are the main ingredients of their decays. The formation of hadrons take place at low energy domain which belongs to the nonperturbative region of QCD. Thus, for the determination of the strong coupling constants of doubly heavy baryons with light mesons, we need some nonperturbative method. At hadronic scale, one should refer to nonperturbative methods in QCD as the strong coupling constant is large and hence perturbative theory becomes invalid. The method of the QCD sum rules \cite{Shifman1979} has proved to be one of the most powerful among all other nonperturbative methods in studying the properties of hadrons. The most advanced version of the method appears to be the light-cone formalism. In the light-cone QCD sum rules (LCSR) (see, for example, \cite{Braun1998}), the operator product expansion (OPE) is performed over twist near the light cone, $ x^2 \sim 0 $. In this case, there appear matrix elements of nonlocal operators between one-particle baryon state and vacuum. These matrix elemenets are parametrized in terms of distribution amplitudes (DAs).
\prg 
Properties of doubly heavy baryons have been studied within the frameworks of lattice QCD \cite{Brown2014}, quark spin symmetry \cite{Hernandez2008, Flynn2012}, and QCD sum rules \cite{Aliev2012, Aliev2013, Azizi2014, Azizi2018, Olamaei2020}. The strong coupling constants of doubly heavy spin-1/2 baryons are examined in \cite{Olamaei2020, Rostami2020, Alrebdi2020, Aliev2020}. The strong coupling constants of light vector mesons with doubly heavy spin-3/2 baryons is estimated in \cite{Azizi2020}.
\prg
This paper is organized as follows. In Sec. \ref{sec:2}, we derive the LCSR for the coupling constants of the light mesons with doubly heavy baryons in the spin-3/2 to spin-1/2 transitions. In Sec. \ref{sec:3}, the numerical analysis of the obtained sum rules is performed. Sec. \ref{sec:4} contains our conclusion.
\section{Light-cone sum rules for the spin-3/2--spin-1/2 doubly heavy baryons with light mesons}\label{sec:2}
To determine the coupling constants of the pseudoscalar and vector mesons with the spin-3/2 to spin-1/2 doubly heavy baryons within the LCSR, we consider the following correlation function:
\begin{align}
	\Pi _\mu = i \int d^4x\ e^{ipx} \langle \mathcal M (q) \vert T\{\eta _\mu (x) \bar \eta (0 )\} \vert 0 \rangle  \label{1}
\end{align}
where $ \mathcal M(q) $ is a light meson with 4-momentum $ q $, and $ \eta _\mu $ and $ \eta $ denote the interpolating current of the corresponding spin-1/2 and spin-3/2 doubly heavy baryons, respectively. The most general form of the interpolating current for spin-1/2 doubly heavy baryons can be written as
\def\Tr{\mathrm{T}}
\begin{gather}
	\eta _\mu = N \epsilon ^{abc} [({q^a}^\Tr B Q^b) Q'^c + ({q^a}^\Tr B Q'^b) Q^c + ({Q^a}^\Tr B Q'^b) q^c]\\
	\eta ^{(S)} = \frac 1{\sqrt 2}\epsilon^{abc} \sum _{i=1}^2 [
		({Q^a}^\Tr A_1^i q^b) A_2^i Q'^c + (Q \leftrightarrow) Q'		
	]\\
	\eta ^{(A)} = \frac 1{\sqrt 6} \epsilon^{abc} \sum _{i=1}^2 [
		2(Q^a A_1^i Q'^b) A_2^i q^c + ({Q^a}^\Tr A_1^i q^b) A_2^i Q'^c - ({Q'^a}^\Tr A_1^i q^b)A_2^i Q^c
	]
\end{gather}
where $ N = \sqrt{1/3}\ (\sqrt{2/3}) $ for identical (different) quarks and
\begin{align}
	A_1^1 = C,\ A_2^1 = \gamma _5, \ A_1^2 = C \gamma _5,\ A_2^2 = \beta I,\ B=C\gamma_\mu
\end{align}
Here, $ \Tr $ is the transpose, $ C $ is the charge conjugation operator, and $ \beta $ is an arbitrary parameter and $ \beta = -1 $ corresponds to the Ioffe current.
\prg 
In order to obtain the LCSR for appropriate quantities, the correlation function is calculated in two different kinematical domains: First, in terms of hadrons, and second, in terms of quark-gluon degrees of freedom in deep Euclidean region by using (OPE). Then, by using the dispersion relation, these two representations are matched and as a result, the desired sum rules are obtained. 
\prg 
The hadronic representation of the correlation function can be obtained by inserting the tower of states carrying the same quantum numbers as the interpolating current and, isolating the ground-state contributions, we obtain
\begin{align}
	\Pi _\mu = \frac{
		\langle 0 \vert \eta _\mu \vert B^* (p_2) \rangle \langle \mathcal M (q) B^* (p_2) \vert B (p_1) \rangle \langle B (p_1) \vert \bar \eta  (0) \vert 0 \rangle
	}{
		(p_2^2-m_2^2) (p_1^2-m_1^2)
	} + \cdots \label{a.2}
\end{align} 
where $ B^*(p_2) $ and $ B(p_1) $ denote the spin-3/2 and spin-1/2 doubly heavy baryons and $ m_2 $ and $ m_1 $ their mass, respectively. In Eq. \eqref{a.2}, $ \cdots $ describes the contribution of higher states and the continuum. 
\prg 
For the calculation of the phenomenological side of the correlation function, the matrix elements $ \langle 0 \vert \eta _\mu  \vert B^*(p_2) \rangle $, $ \langle \mathcal M (q) B^*(p_2) \vert B (p_1) \rangle $, and $ \langle B(p_1) \vert \eta  \vert 0 \rangle $ are needed. These matrix elements are determined in the following way:
\begin{gather}
	\langle 0 \vert \eta \vert B(p_1) \rangle = \lambda _1 u(p_1) \label{a.3}\\
	\langle 0 \vert \eta _\mu \vert B^* (p_2) \rangle = \lambda _2  u _\mu (p_2) \label{a.4}\\
	\langle \mathcal P (q) B^* (p_2) \vert B (p_1) \rangle = g \bar u _\alpha (p_2) u (p_1)q^\alpha \label{a.5}
\end{gather}
The matrix element $ \langle  V (q) B(p_2) \vert B^* (p_1) \rangle $ is parametrized in terms of three couplings as follows \cite{Jones1973}:
\begin{align}
	\langle V (q)  B^*(p_2) \vert B (p_1) \rangle &= \bar u _\alpha (p_2) [
		g_1 (q_\alpha \slashed \varepsilon - \varepsilon _\alpha \slashed q) \gamma _5
		- g_2 (P\cdot \varepsilon q_\alpha - P\cdot q \varepsilon _\alpha) \gamma _5
		\nn + g_3 (q\cdot \varepsilon q_\alpha - q^2 \varepsilon _\alpha) \gamma _5 
	] u (p_1) \label{a.6}
\end{align}
where $ u(p_2) $ is the Dirac bispinor for spin-1/2 baryons whilst $ u_\alpha (p_1) $ is the Rarita-Schwinger spinor for spin-3/2 baryons, $ \varepsilon_\mu $ is the polarization 4-vector of the light vector meson, $ P = (p_1+p_2)/ 2 $, and $ q = p_1-p_2 $. In the following discussions, we denote $ p_2=p $ and $ p_1=p+q $ as well as we impose the on-shell condition for the vector meson, namely $ q^2 = m_{ V}^2 $ and consequently $ q\cdot \varepsilon = 0 $.
\prg 
Taking into account Eqs. \eqref{a.3}--\eqref{a.6} in Eq. \eqref{a.2} and using the completeness conditions for Dirac and Rarita-Schwinger spinors which reads
\begin{gather}
	\sum _s u (p) \bar u (p) = \slashed p + m \label{a.7}\\
	\sum _s u _\alpha (p) \bar u _\beta (p) = - (\slashed p + m) \pp{g _{\alpha\beta} - \frac 13 \gamma _\alpha \gamma _\beta + \frac 23 \frac{p_\alpha p_\beta}{m^2} + \frac{p_\alpha \gamma _\beta - p_\beta \gamma _\alpha}{3m}} \label{a.8}
\end{gather}
one can easily get the expressions for the strong coupling constants of the light mesons with the doubly heavy baryons. Before moving to the next calculations, here we would like to bring the attention of the reader to the existence of two problems:
\begin{enumerate}
	[i)]
	\item The negative-parity spin-1/2 baryons contribute to the matrix element $ \langle 0 \vert \eta _\mu \vert B (p) \rangle $. From general considerations, this matrix element can be parametrized as
	\begin{align}
		\langle 0 \vert \eta _\mu \vert B (p_1) \rangle \sim (\gamma _\mu - A p _{1,\mu}) u (p_1,s) \label{a.9}
	\end{align}
	From this equation, it follows that the structures proportional to $ \gamma _\mu $ and $ p_{1,\mu} $ contain the contributions coming not only from spin-3/2 baryons, but also from spin-1/2 baryons, which should be eliminated. Therefore, we will discard these structures in the next discussions. From Eq. \eqref{a.8}, it follows that only the structures proportional to $ g _{\alpha\beta} $ contain solely the contribution of the spin-3/2 state.
	\item Not all Lorentz structures are independent. This problem can be overcome by using the specific order of Dirac matrices. For the calculations of the strong coupling constants of pseudoscalar (vector) mesons with the spin-3/2 and spin-1/2 baryons, the ordered Dirac matrices are chosen in the form $ \slashed q \slashed p \gamma _\mu \ (\gamma _\mu \slashed \varepsilon \slashed q \slashed p \gamma _5) $.
\end{enumerate}
Taking into account these facts and using Eqs. \eqref{a.3}--\eqref{a.6} in the phenomenological parts of the correlation functions, we finally get
\begin{align}
	\Pi _\mu ^{(\mathcal P)} = \frac{
		g\lambda _1 \lambda _2 \slashed q \slashed p q_\mu 
	}{
		 (m_2^2-p^2)[m_1^2 - (p+q)^2]
	} + {\rm other\ structures }\label{a.99}
\end{align}
and
\begin{align}
	\Pi _\mu ^{(\mathcal V)} &=  \frac{\lambda_1\lambda_2}{[m_1^2 - (p+q)^2][m_2^2 - p^2]}[
		g_1 (m_1+m_2)\slashed \varepsilon \slashed p \gamma _5 q _\mu 
		+ g _2 \slashed q \slashed p \gamma _5 p \cdot \varepsilon q _\mu 
		- g_3 m_V^2 \slashed q \slashed p \gamma _5 \varepsilon _\mu 
		\nn + {\rm\ other\ structures}
	] \label{a.10}
\end{align}
To obtain the LCSR for the aforementioned coupling constants, we need to calculate the correlation function from the QCD side and choose the coefficients of the same structures and then match with the results from the hadronic part. The expression of the correlation function is obtained by using OPE at deep Euclidean region, $ -p^2 \to \infty $ and $ -(p+q)^2 \to \infty $. Using the Wick theorem, from Eq. \eqref1, one can get
\begin{align}
	\Pi _\mu ^{(S)} &= \sqrt{\frac 13} \epsilon^{abc} \epsilon^{a'b'c'} \int d^4x\ e^{ipx} \sum _{i=1}^2 B _{\alpha\beta} (\tilde A_2^i ) _{\gamma'\rho'} (\tilde A_1^i) _{\alpha'\beta'}\langle \mathcal M (q) \vert (
			-S_{Q'\gamma\gamma'}^{cc'}S_{Q\beta\beta'}^{ba'}q_\alpha^{a}\bar q_{\alpha'}^{b'}
			-S_{Q\gamma\gamma'}^{cc'} S_{Q'\beta\beta'}^{ba'} q_\alpha^a \bar q_{\alpha'}^{b'}
			\nn +S_{Q\beta\gamma'}^{bc'} S_{Q'\gamma\beta'}^{ca'} q_\alpha^a \bar q_{\alpha'}^{b'}
			+S_{Q'\beta\gamma'}^{bc'} S_{Q\gamma\beta'}^{ca'} q_\alpha^a \bar q_{\alpha'}^{b'}
			-S_{Q'\beta\gamma'}^{bc'} S_{Q\alpha\beta'} q_\gamma^c \bar q _{\alpha'}^{b'}
			+S_{Q\alpha\gamma'}^{ac'} S_{Q'\beta\beta'}^{ba'} q_\gamma^c \bar q _{\alpha'}^{b'}
		) \vert 0 \rangle \label{a.11}
\end{align}
and
\begin{align}
	\Pi _\mu ^{(A)} &= \sqrt{\frac 13} \epsilon^{abc} \epsilon^{a'b'c'} \int d^4x\ e^{ipx} \sum _{i=1}^2 B _{\alpha\beta} (\tilde A _2^i) _{\gamma'\rho'} (\tilde A _1^i) _{\alpha'\beta'} \langle \mathcal M (q) \vert (
		-S_{Q'\gamma\gamma'}^{cc'} S_{Q\beta\beta'}^{ba'} q_\alpha^a \bar q_{\alpha'}^{b'}
		+S_{Q'\beta\beta'}^{ba'} S_{Q\gamma\gamma'}^{cc'} q_\alpha^a \bar q_{\alpha'}^{b'}
		\nn +2 S_{Q'\beta\alpha'}^{bb'} S_{Q\alpha\beta'}^{aa'} q_\gamma^c \bar q_{\gamma'}^{c'}
		+2 S_{Q'\gamma\alpha'}^{cb'} S_{Q\beta\beta'}^{ba'} q_\alpha^a \bar q_{\gamma'}^{c'}
		-2 S_{Q'\beta\alpha'}^{bb'} S_{Q\gamma\beta'}^{ca'} q_\alpha^a \bar q_{\gamma'}^{c'}
		-S_{Q\beta\gamma'}^{bc'} S_{Q'\gamma\beta'}^{ca'} q_\alpha^a \bar q_{\alpha'}^{b'}
		\nn +S_{Q'\beta\gamma'}^{bc'} S_{Q\gamma\beta'}^{ca'} q_\alpha^a \bar q_{\alpha'}^{b'}
		-S_{Q'\beta\gamma'}^{bc'} S_{Q\alpha\beta'}^{aa'} q_\gamma^c \bar q_{\alpha'}^{b'}
		-S_{Q\alpha\gamma'}^{ac'} S_{Q'\beta\beta'}^{ba'} q_\gamma^c \bar q_{\alpha'}^{b'}
	) \vert 0 \rangle \label{a.12}
\end{align}
Here, $ S_Q $ is the heavy quark propagator. The superscript $ (S) $ and $ (A) $ denote the symmetric and antisymmetric currents of spin-1/2 baryons and $ \tilde A _k^i = \gamma ^0 {A_k^i}^\dagger \gamma^0 $. The heavy quark propagator in the presence of an external background field is given in the coordinate representation by
\begin{align}
	S_{Q\alpha\beta}^{aa'} (x) &= \frac{m_Q^2}{4\pi} \Big[ 
		\frac{iK_2(m_Q \sqrt{-x^2})}{(\sqrt{-x^2})^2} 
		+ \frac{m_Q^2 K_1 (m_Q \sqrt{-x^2} ) } {\sqrt{-x^2}}
	\Big] _{\alpha\beta} \delta ^{aa'} 
	- \frac{g_s}{16\pi^2} m_Q \int_0^1 du \ \nn\times \Big[ 
		\frac{iK_1 (m_Q\sqrt{-x^2})  } {\sqrt{-x^2}} (u\slashed x \sigma _{\lambda\tau} + \bar u \sigma _{\lambda\tau} \slashed x) + K_0 (m_Q \sqrt{-x^2}) \sigma _{\lambda \tau}
	\Big] _{\alpha\beta} G^{(n)\lambda\tau} \pp{\frac{\lambda^{n}}2}^{aa'} \label{a.13}
\end{align}
where $ G^{(n)} _{\lambda\tau} $ is the gluon field strength tensor, the $ \lambda ^{(n)} $ are the Gell-Mann matrices, and the $ K_i (m_Q \sqrt{-x^2}) $ are the modified Bessel functions of the second kind. 
\prg 
Using the Fiertz identities
\begin{gather}
	q_\alpha^b \bar q_\beta ^{b'} = -\frac 1{12} (\Gamma _i)_{\alpha\beta} \delta ^{bb'} \bar q \Gamma _i q \label{a.14}\\
	q_\alpha^b \bar q_\beta^{b'} G _{\lambda\tau}^{(n)} = -\frac 1{16} \pp{\frac{\lambda ^{(n)}}2}^{bb'} (\Gamma _i)_{\alpha\beta} \bar q \Gamma _i G _{\lambda \tau}^{(n)} q \label{a.15}
\end{gather}
from Eqs. \eqref{a.11} and \eqref{a.12}, it follows that the problem for the calculation of the correlation functions from the QCD side reduces to the determination of the matrix elements $ \langle \mathcal M (q) \vert \bar q \Gamma _i q \vert 0 \rangle $ and $ \langle \mathcal M (q) \vert \bar q \Gamma _i G _{\lambda\tau}^{(n)} q \vert 0 \rangle $, where $ \{\Gamma _i\} $ is the full set of Dirac matrices,
\begin{align}
	\Gamma _i = \cc{I, \gamma_5, \gamma_\mu, i\gamma_\mu\gamma_5, \frac{1}{\sqrt 2}\sigma _{\mu\nu}}_{i=1}^5
\end{align}
These matrix elements are the main nonperturbative input parameters of the LCSR and they are expressed in terms of light meson DAs of different twists. The expressions of these matrix elements in terms of meson DAs are found in \cite{Ball1998, Ball1999, Ball1996, Ball2006, Ball19992, Ball2005} and for completeness, we present their expressions in Appendix \ref{app:A}.
\prg 
Inserting Eqs. \eqref{a.14} and \eqref{a.15} into \eqref{a.11} and \eqref{a.12}, performing double Borel transformation over the variables $ -p^2 $ and $ -(p+q)^2 $ in both representations of the correlation function, and choosing the coefficients of the corresponding structures, we get the following sum rules:
\begin{gather}
	g = \frac{1}{\lambda _1 \lambda _2} e^{m_1^2/M_1^2+m_2^2/M_2^2} \Pi ^{{\rm theo} (S,A)} \label{a.16}\\
	g _1 = \frac{1}{\lambda _1 \lambda _2 (m_1+m_2)} e^{m_1^2/M_1^2+m_2^2/M_2^2} \Pi _1^{{\rm theo} (S,A)} \label{a.17}\\
	g _2 = \frac{1}{\lambda _1 \lambda _2} e^{m_1^2/M_1^2+m_2^2/M_2^2} \Pi _2^{{\rm theo} (S,A)} \label{a.17.2}\\
	g _3 = -\frac{1}{\lambda _1 \lambda _2 m_V^2} e^{m_1^2/M_1^2+m_2^2/M_2^2} \Pi _3^{{\rm theo} (S,A)} \label{a.17.3}
\end{gather}
where $ \Pi ^{{\rm theo}(A)} = 0 = \Pi _i^{{\rm theo}(A)} $ for all the coupling constants and
\begin{align}
	\Pi ^{{\rm theo}(S)} &= \frac{im_Q^2 m_{Q'}^2}{12 \sqrt 3\pi^4 M^4} \{
		6(-1+\beta) f_{\mathcal P} \mathcal I _{22}[\phi _{\mathcal P}(u)]
		-(1+\beta) \mu _{\mathcal P} (-1+\tilde \mu _{\mathcal P}^2) \mathcal I_{12} [\phi _\sigma (u)]
	\}\\
	\Pi _1^{{\rm theo}(S)} &= -\frac{im_Q^2 m_{Q'}^2 }{32\sqrt 3\pi^4 M^4} \{
		16(-1+\beta) f_{V} m _{ V} \mathcal I_{22}[\phi _3^\perp (u)]
		+ f _{ V}^T [
			16M^2(1+2\beta) \mathcal I_{12} [\phi _2^\perp (u)] \nn + m _{ V}^2 (8i(1+\beta) \mathcal I_{12}[\Hat [1,\phi _2^\perp (u)] - \Hat[1, \psi _4^\perp (u)] )
			+ M^2 (1+2\beta) I _{12}^2 [\phi _4^\perp (u)]
		]
	\}\\
	\Pi _2^{{\rm theo}(S)} &= \frac{m_Q^2 m_{Q'}^2 }{2\sqrt 3\pi^4 M^6} \{
		i(1+\beta) f_{V}^T \mathcal I _{12} [\Hat [1,\phi _2^\perp (u)] - 2 \Hat[2, \phi _3^\perp (u)] + \Hat[2,\psi  _4^\perp (u)]]
		\nn + 2(-1+\beta) f _{ V} m _{ V} \mathcal I _{22} [\Hat[1, \phi _2^\parallel (u)] - \Hat [1,\phi _3 ^\perp (u)]]
	\}\\
	\Pi _3 ^{{\rm theo}(S)} &= -\frac{im_Q^2 m_{Q'}^2}{32\sqrt 3\pi^4 M^4} \{
		32 i (-1+\beta) f _{ V} m _{ V} \mathcal I _{22} [\Hat[1, \phi_2^\parallel (u) ] - \Hat[1, \phi _3^\perp (u)] ]
		\nn + f _{ V} ^T (
			-16(1+\beta) \mathcal I _{12}[\Hat[2,  \phi _2^\perp (u)] - 2 \Hat[2, \phi _3 ^\parallel (u) ] + \Hat[2,  \psi _4^\perp (u)] ] 
			\nn + 16(1+2\beta) M^2 \mathcal I _{12} [\Hat[1, \phi _2^\perp (u)] ]
			+ (1+2\beta) m _{ V}^2 M^2 I _{12}[\phi _4^\perp (u)]
		)
	\}
\end{align}
where we have defined the integrals and operators
\ba 
\I _{ij} [f(u)] := \int du \int d^4x\ e^{i(p+\bar uq)\cdot x} K_i K_j f(u) 
\ea 
\ba 
\I _{ij}^2 [f(u)] := \int du \int d^4x\ e^{i(p+\bar uq)\cdot x} K_i K_j f(u) x^2
\ea 
\ba 
\I _{ij}^4 [f(u)] := \int du \int d^4x\ e^{i(p+\bar uq)\cdot x} K_i K_j f(u) x^4
\ea 
\ba 
\Hat [n,f(u)] := i^n \int _0^u dv_n \cdots \int _0^{v_3} dv_2 \int _0^{v_2} dv_1\ f(v_1)
\ea 
and we have introduced the short-hand notation 
\ba 
K _i := \frac{K_i(m_Q\sqrt{-x^2})}{(\sqrt{-x^2})^i}, \quad K _j := \frac{K_j (m_{Q'}\sqrt{-x^2})}{(\sqrt{-x^2})^j}
\ea 
Here, for the sake of simplicity, we omit the contributions coming from the matrix elements of 3-particle nonlocal operators  between the vacuum and 1-particle meson state but we include them in our numerical analysis. The details of the calculations of the correlation function from the QCD side is presented in Appendix \ref{app:B}.
\section{Numerical analysis}\label{sec:3}
\begin{figure}
	[H]\centering
	\includegraphics[width=\textwidth]{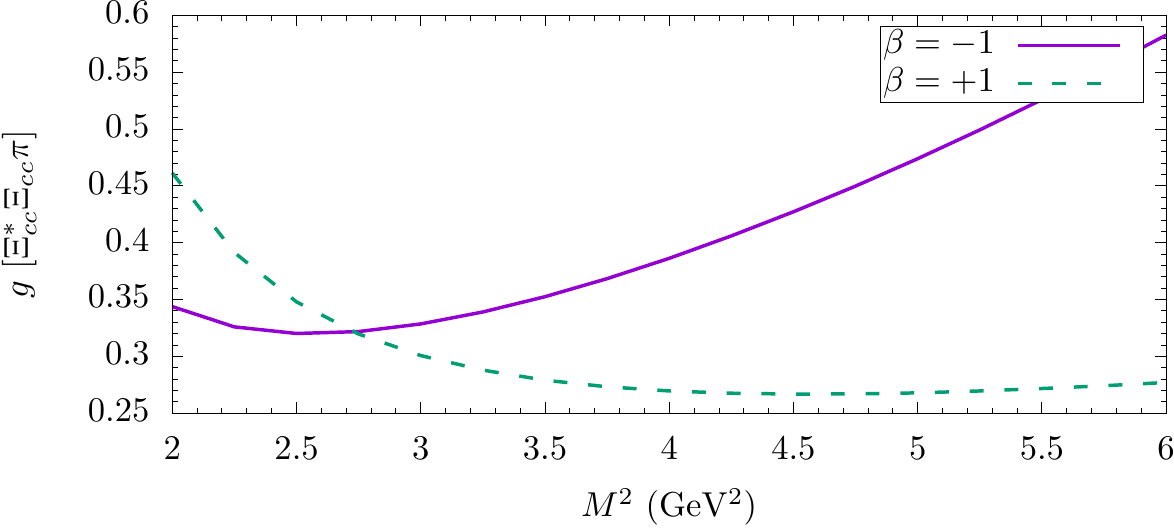}
	\caption{The dependence of the coupling constant $ g $ of the pion in the transition $ \Xi _{cc}^* $ to $ \Xi _{cc} $ on $ M^2 $ at shown values of $ \beta $ and the fixed value of $ s_0 = 18 {\rm\ GeV^2} $.}\label{fig:1}
\end{figure}

\begin{figure}
	[H]\centering
	\includegraphics[width=\textwidth]{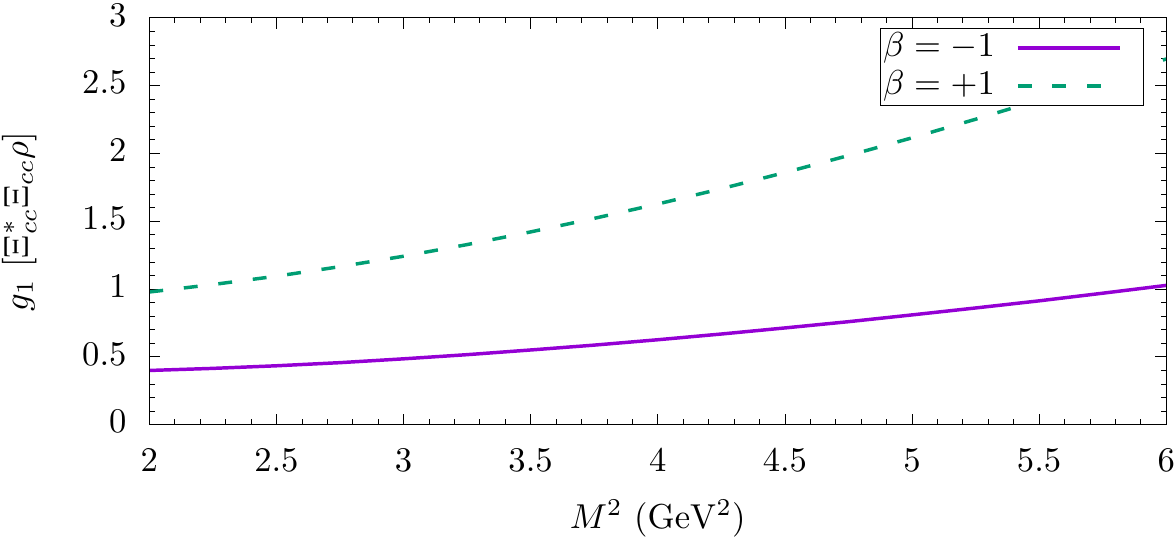}
	\caption{The dependence of the coupling constant $ g_1 $ of the $ \rho $ meson in the transition $ \Xi _{cc}^* $ to $ \Xi _{cc} $ on $ M^2 $ at shown values of $ \beta $ and the fixed value of $ s_0 = 18 {\rm\ GeV^2} $.}\label{fig:2}
\end{figure}

\begin{figure}
	[H]\centering
	\includegraphics[width=\textwidth]{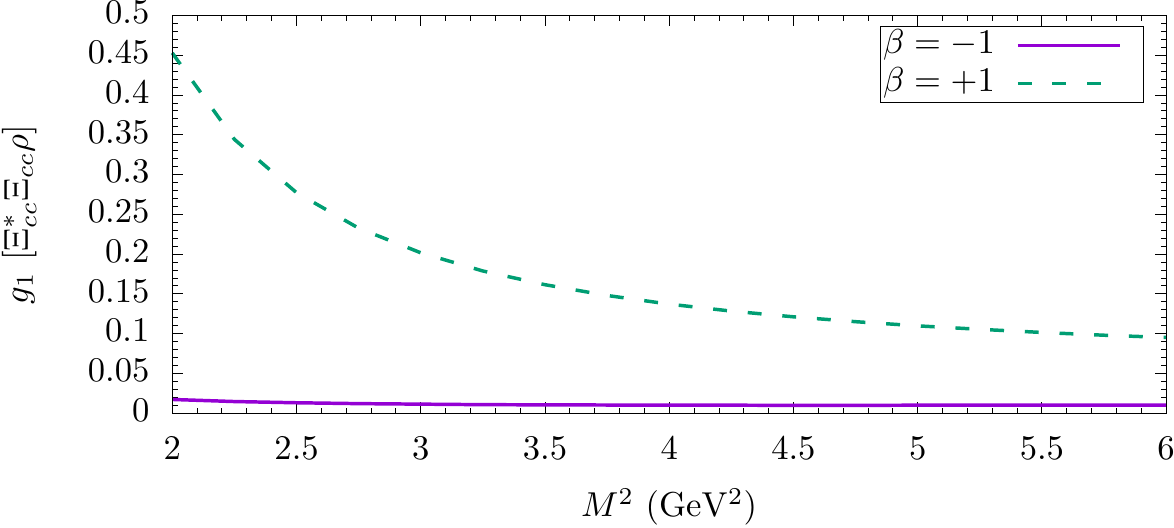}
	\caption{The dependence of the coupling constant $ g_1 $ of the $ \rho $ meson in the transition $ \Xi _{cc}^* $ to $ \Xi _{cc} $ on $ M^2 $ at shown values of $ \beta $ and the fixed value of $ s_0 = 18 {\rm\ GeV^2} $.}\label{fig:3}
\end{figure}

\begin{figure}
	[H]\centering
	\includegraphics[width=\textwidth]{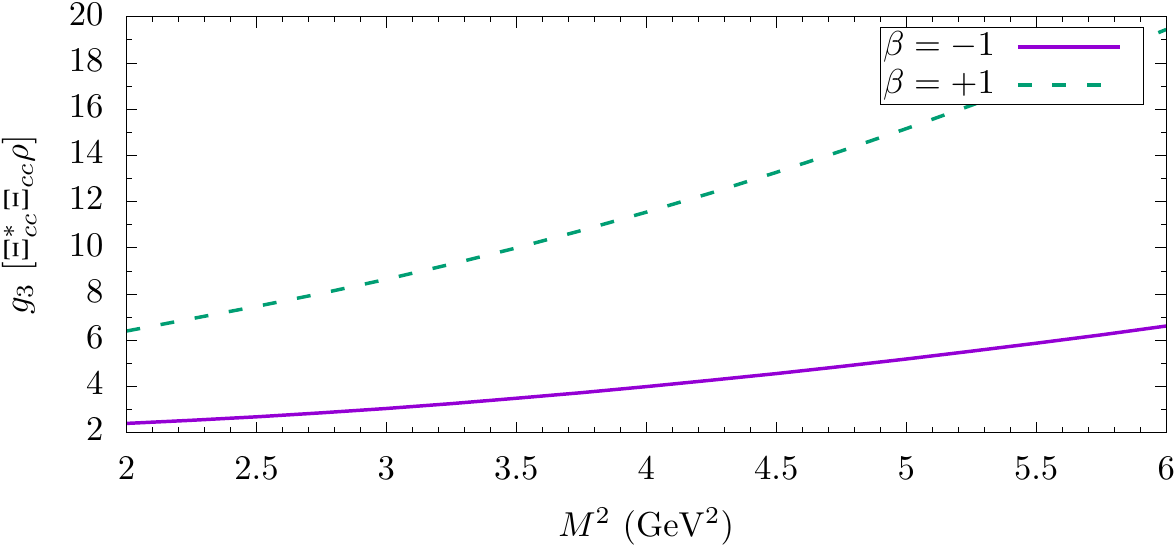}
	\caption{The dependence of the coupling constant $ g_1 $ of the $ \rho $ meson in the transition $ \Xi _{cc}^* $ to $ \Xi _{cc} $ on $ M^2 $ at shown values of $ \beta $ and the fixed value of $ s_0 = 18 {\rm\ GeV^2} $.}\label{fig:4}
\end{figure}

\begin{figure}
	[H]\centering
	\includegraphics[width=\textwidth]{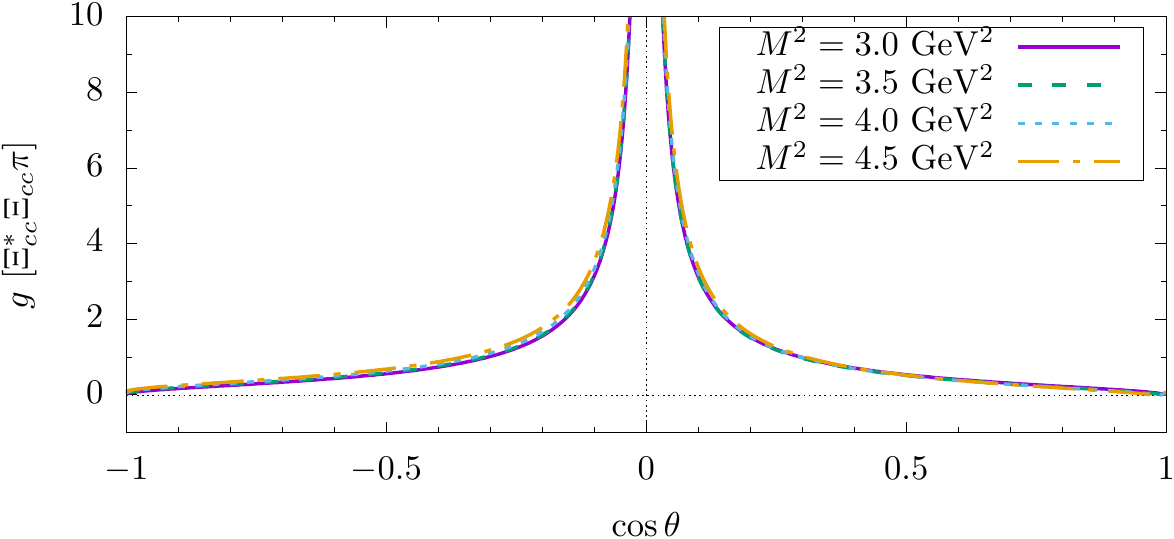}
	\caption{The dependence of the coupling constant $ g $ of the pion in the transition $ \Xi _{cc}^* $ to $ \Xi _{cc} $ on $ \cos\theta $ at shown values of $ M^2 $ and the fixed value of $ s_0 = 18 {\rm\ GeV^2} $.}\label{fig:5}
\end{figure}

\begin{figure}
	[H]\centering
	\includegraphics[width=\textwidth]{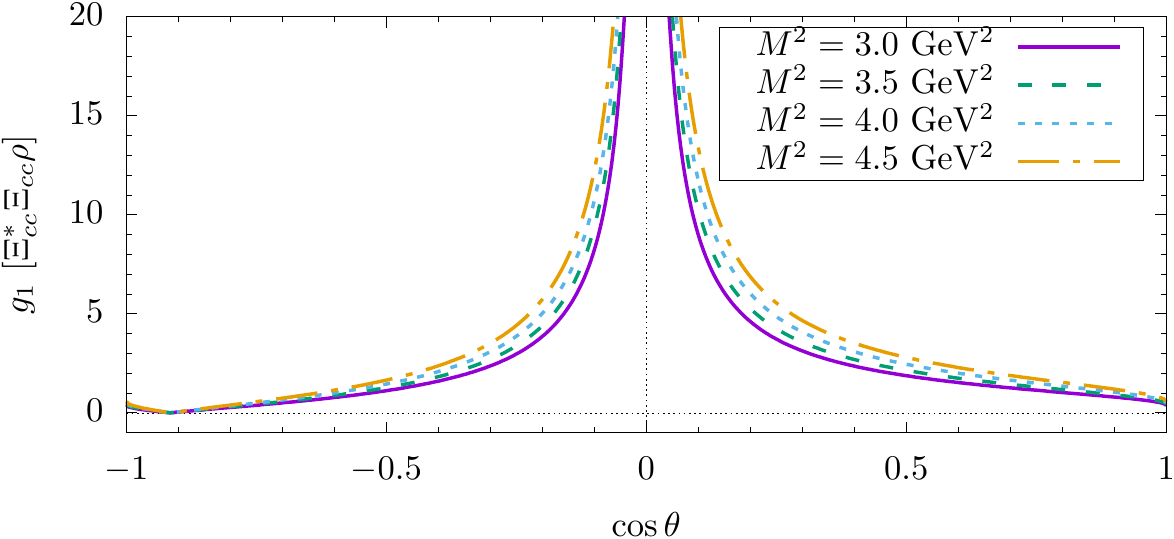}
	\caption{The dependence of the coupling constant $ g_1 $ of the $ \rho $ meson in the transition $ \Xi _{cc}^* $ to $ \Xi _{cc} $ on $ \cos\theta $ at shown values of $ M^2 $ and the fixed value of $ s_0 = 18 {\rm\ GeV^2} $.}\label{fig:6}
\end{figure}

\begin{figure}
	[H]\centering
	\includegraphics[width=\textwidth]{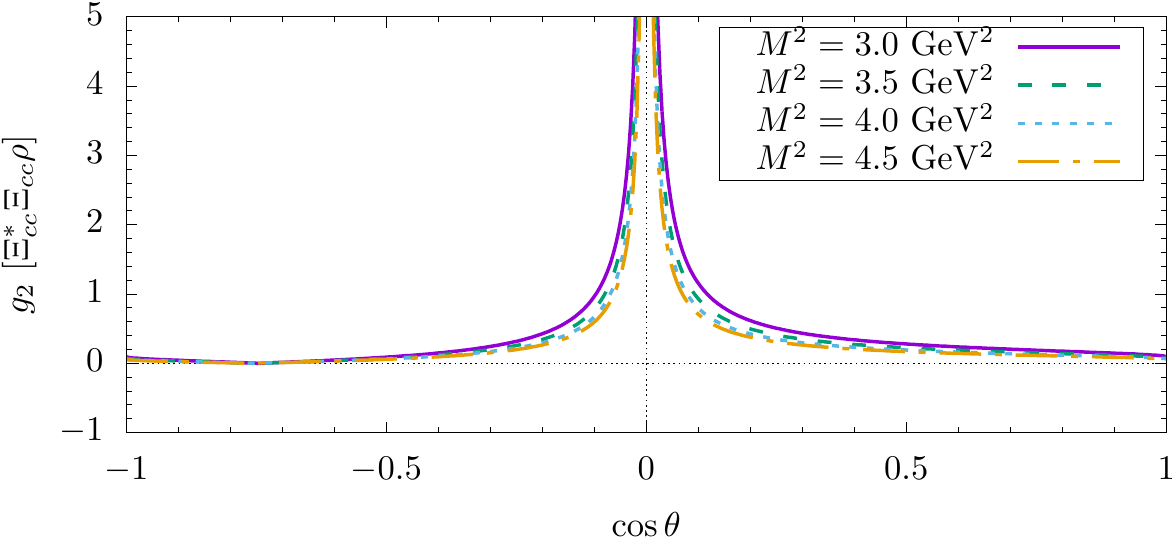}
	\caption{The dependence of the coupling constant $ g_2 $ of the $ \rho $ meson in the transition $ \Xi _{cc}^* $ to $ \Xi _{cc} $ on $ \cos\theta $ at shown values of $ M^2 $ and the fixed value of $ s_0 = 18 {\rm\ GeV^2} $.}\label{fig:7}
\end{figure}

\begin{figure}
	[H]\centering
	\includegraphics[width=\textwidth]{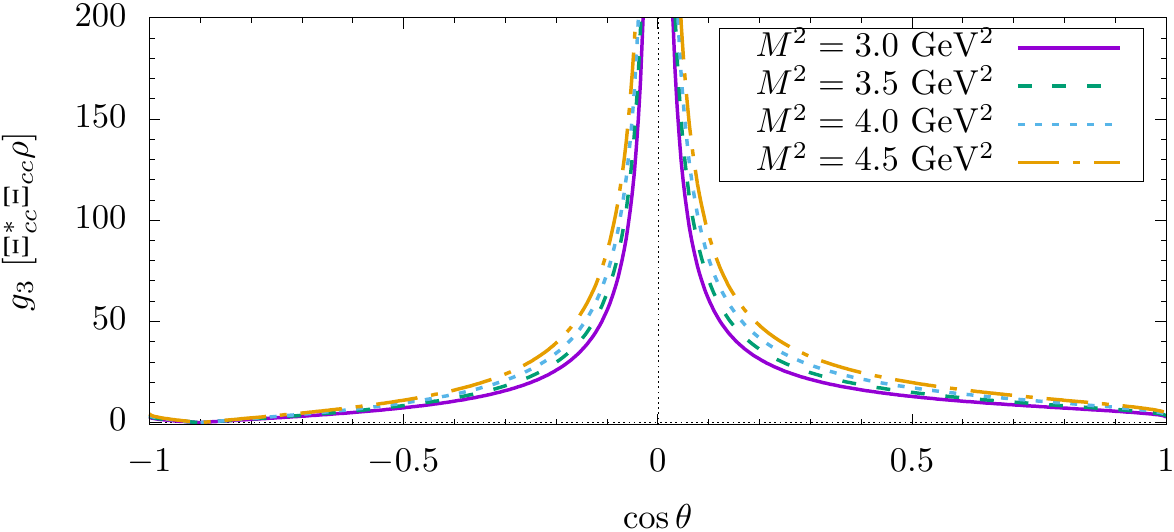}
	\caption{The dependence of the coupling constant $ g_3 $ of the $ \rho $ meson in the transition $ \Xi _{cc}^* $ to $ \Xi _{cc} $ on $ \cos\theta $ at shown values of $ M^2 $ and the fixed value of $ s_0 = 18 {\rm\ GeV^2} $.}\label{fig:8}
\end{figure}

In this section, we numerically analyze the LCSR for the coupling constants $ g $ and $ g _i $ of the light mesons $ \pi $, $ K $, $ \rho $, and $ K^* $ in the transition of the spin-3/2 to spin-1/2 doubly heavy baryons, namely $ \Xi _{cc}^* $, $ \Xi _{bb}^* $, $ \Xi _{bc}^* $, $ \Omega _{cc}^* $, $ \Omega _{bb}^* $, $ \Omega _{bc}^* $ to $ \Xi _{cc} $, $ \Xi _{bb} $, $ \Xi _{bc} $, $ \Omega _{cc} $, $ \Omega _{bb} $, $ \Omega _{bc} $, by using Package X \cite{Patel2015}. The LCSR for the coupling constants $ g $, $ g _1 $, $ g _2 $, and $ g _3 $ involve various input parameters such as quark masses, the masses and decay constants of the light mesons, and the masses and residues of the said doubly heavy baryons. Some of these parameters are presented in Table \ref{tab:1}. Another set of essential input parameters are meson DAs of different twists, which are given in Appendix \ref{app:A}.
{
\renewcommand{\arraystretch}{1.2}
\def\para{Parameter}
\def\val{Value}
\def\ms{$ m_s {\rm\ (1\ GeV)} $}
\def\mc{$ m_c $}
\def\mb{$ m_b $}
\def\mrho{$ m_\rho $}
\def\frho{$ f_\rho $}
\def\frhoT{$ f_\rho^T $}
\def\mks{$ m_{K^*} $}
\def\fks{$ f_{K^*} $}
\def\fksT{$ f_{K^*}^T $}
\def\mpi{$ m_\pi $}
\def\fpi{$ f_\pi $}
\def\mk{$ m_K $}
\def\fk{$ f_K $}
\def\mxicc{$ m_{\Xi_{cc}} $}
\def\mxibb{$ m_{\Xi_{bb}} $}
\def\mxibc{$ m_{\Xi_{bc}} $}
\def\momcc{$ m_{\Omega_{cc}} $}
\def\mombb{$ m_{\Omega_{bb}} $}
\def\mombc{$ m_{\Omega_{bc}} $}
\def\lxicc{$ \lambda_{\Xi_{cc}} $}
\def\lxibb{$ \lambda_{\Xi_{bb}} $}
\def\lxibc{$ \lambda_{\Xi_{bc}} $}
\def\lomcc{$ \lambda_{\Omega_{cc}} $}
\def\lombb{$ \lambda_{\Omega_{bb}} $}
\def\lombc{$ \lambda_{\Omega_{bc}} $}
\def\mxiccs{$ m_{\Xi_{cc}^*} $}
\def\mxibbs{$ m_{\Xi_{bb}^*} $}
\def\mxibcs{$ m_{\Xi_{bc}^*} $}
\def\momccs{$ m_{\Omega_{cc}^*} $}
\def\mombbs{$ m_{\Omega_{bb}^*} $}
\def\mombcs{$ m_{\Omega_{bc}^*} $}
\def\lxiccs{$ \lambda_{\Xi_{cc}^*} $}
\def\lxibbs{$ \lambda_{\Xi_{bb}^*} $}
\def\lxibcs{$ \lambda_{\Xi_{bc}^*} $}
\def\lomccs{$ \lambda_{\Omega_{cc}^*} $}
\def\lombbs{$ \lambda_{\Omega_{bb}^*} $}
\def\lombcs{$ \lambda_{\Omega_{bc}^*} $}
\begin{table}
	[H]\centering
	\caption{Some of the values of the input parameters entering the sum rules. All the masses and decay constants are in units of GeV.}\label{tab:1}
	\begin{tabular}
		{|cc|cc|cc|}
		\hline
		\hline 
		\para	& \val	& \para	 					& \val	& \para	 & \val\\
		\hline 
		\ms		& 0.137	& \mxicc\ \cite{Aliev2012} 	& 3.72	& \lxicc\ \cite{Aliev2012} & 0.16\\
		\mc		& 1.4	& \mxibb\ \cite{Aliev2012} 	& 9.96	& \lxibb\ \cite{Aliev2012} & 0.44\\
		\mb		& 4.7	& \mxibc\ \cite{Aliev2012} 	& 6.72	& \lxibc\ \cite{Aliev2012} & 0.28\\
		\mrho	& 0.770	& \momcc\ \cite{Aliev2012} 	& 3.73	& \lomcc\ \cite{Aliev2012} & 0.18\\
		\frho	& 0.216	& \mombb\ \cite{Aliev2012} 	& 9.97	& \lombb\ \cite{Aliev2012} & 0.45\\
		\frhoT	& 0.165	& \mombc\ \cite{Aliev2012} 	& 6.75	& \lombc\ \cite{Aliev2012} & 0.29\\
		\mks	& 0.892 & \mxiccs\ \cite{Aliev2013}	& 3.69  & \lxiccs\ \cite{Aliev2013}& 0.12\\
		\fks	& 0.220 & \mxibbs\ \cite{Aliev2013} & 10.4  & \lxibbs\ \cite{Aliev2013}& 0.22\\
		\fksT	& 0.185 & \mxibcs\ \cite{Aliev2013}	& 7.25  & \lxibcs\ \cite{Aliev2013}& 0.15\\
		\mpi	& 0.135 & \momccs\ \cite{Aliev2013}	& 3.78  & \lomccs\ \cite{Aliev2013}& 0.14\\
		\fpi	& 0.131 & \mombbs\ \cite{Aliev2013}	& 10.5  & \lombbs\ \cite{Aliev2013}& 0.25\\
		\mk		& 0.495 & \mombcs\ \cite{Aliev2013}	& 7.3   & \lombcs\ \cite{Aliev2013}& 0.17\\
		\fk		& 0.160 &		 &		&		 & \\
		\hline 
		\hline 
	\end{tabular}
\end{table}
}
In addition to the input parameters summarized in Table \ref{tab:1} and the meson DAs, the LCSR involves three auxiliary parameters, i.e. the Borel mass parameter, $ M^2 $, the continuum threshold, $ s_0 $, and the arbitrary parameter, $ \beta $, which appears in the interpolating current. Hence, one should find the working regions of these parameters so that the LCSR is reliable. The lower bound of the Borel mass parameter can be obtained by insisting on that the contributions from the highest-twist terms should be considerably smaller than the contributions from the lowest-twist terms. On the other hand, the upper limit of $ M^2 $ can be determined by demanding that the continuum contribution should not be too large. Meanwhile, the continuum threshold, $ s _0 $, is obtained by requiring that the two-point sum rules reproduce a 10\% accuracy of the mass of the doubly heavy baryons. These criteria lead to the values of $ M^2 $ and $ s_0 $ summarized in Table \ref{tab:2} for the transitions considered.
{
\renewcommand{\arraystretch}{1.2}
\begin{table}
	[H]\centering 
	\caption{The working region of the parameters $ M^2 $ and $ s_0 $ for the transitions considered in our work. Here, $ \mathcal M _1 = \pi {\rm\ or\ } \rho $ and $ \mathcal M _2 = K {\rm\ or\ } K^* $.}\label{tab:2}
	\begin{tabular}
		{|c|c|c||c|c|c|}
		\hline 
		\hline 
		Channel 						   			& $ M^2 {\rm\ (GeV^2)} $ 	& $ s_0 {\rm\ (GeV^2)} $ \\
		\hline 
		$ \Xi_{cc}^* \to \Xi_{cc} \mathcal M_1 $    & $ 3.0 \leq M^2 \leq 4.5 $	& $ 18\pm 4 $ \\
		$ \Xi_{bb}^* \to \Xi_{bb} \mathcal M_1 $    & $ 8 \leq M^2 \leq 12 $	& $ 110\pm 10 $ \\
		$ \Xi_{bc}^* \to \Xi_{bc} \mathcal M_1 $   	& $ 6 \leq M^2 \leq 8 $		& $ 60\pm 5 $ \\
		$ \Omega_{cc}^* \to \Xi_{cc} \mathcal M_2 $ & $ 3.0 \leq M^2 \leq 4.5 $	& $ 18\pm 4 $ \\
		$ \Omega_{bb}^* \to \Xi_{bb} \mathcal M_2 $ & $ 8 \leq M^2 \leq 12 $	& $ 110\pm 10 $ \\
		$ \Omega_{bc}^* \to \Xi_{bc} \mathcal M_2 $ & $ 6 \leq M^2 \leq 8 $		& $ 60\pm 5 $ \\
		\hline 
		\hline 
	\end{tabular}
\end{table}
}
Our analysis reveals that the contributions from the twist-4 terms in the considered domains of $ M^2 $ at the shown values of $ s_0 $ are smaller than 17\% and higher states contribute 28\% at maximum for all the considered channels. As an example, we present the $ M^2 $ dependence of $ g $, $ g_1 $, $ g_2 $, and $ g_3 $ for $ \Xi_{cc}^*\to \Xi_{cc}\pi $ and $ \Xi_{cc}^*\to\Xi_{cc}\rho $ at fixed values of $ s_0 $ and $ \beta $ in Figs. \ref{fig:1}--\ref{fig:4}, respectively. Having determined the working regions of $ M^2 $ and $ s_0 $, one should find the working region of the auxiliary parameter, $ \beta $. To do so, we investigate the dependence of $ g $, $ g_1 $, $ g_2 $, and $ g_3 $ on $ \cos\theta $, where $ \theta $ is defined through the relation $ \beta = \tan \theta $. As an illustration, we present the dependence of the said coupling constants for the transition $ \Xi_{cc}^*\to\Xi_{cc} $ at fixed values of $ M^2 $ and $ s_0 $ in Figs. \ref{fig:5}--\ref{fig:8}, respectively.
In Figs. \ref{fig:5}--\ref{fig:8}, we observe that the coupling constants remain practically the same when $ \vv{\cos\theta} $ varies between 0.5 and 0.8. Our numerical analysis for the coupling constants of the doubly heavy baryons with the light mesons leads to the results presented in Tables \ref{tab:3} and \ref{tab:4}. The uncertainties are due to the variation of the parameters $ M^2 $, $ s_0 $, and the errors in the values of the input parameters.
{
\renewcommand{\arraystretch}{1.2}
\begin{table}
	[H]\centering 
	\caption{The numerical values of the coupling constants, $ g $, of the light pseudoscalar mesons with the doubly heavy baryons.}\label{tab:3}
	\begin{tabular}
		{|c||c|c|}
		\hline 
		\hline 
		Channel & Case of the general current & Case of the Ioffe current \\
		\hline 
		$ \Xi_{cc}^* \to \Xi_{cc} \pi $ & $ 0.39 \pm 0.02 $ & $ 0.37 \pm 0.04 $\\
		$ \Xi_{bb}^* \to \Xi_{bb} \pi $ & $ 0.22 \pm 0.03 $ & $ 0.20 \pm 0.04 $\\
		$ \Xi_{bc}^* \to \Xi_{bc} \pi $ & $ 0.30 \pm 0.02 $ & $ 0.26 \pm 0.3 $\\
		$ \Omega_{cc}^* \to \Xi_{cc} K $ & $ 0.99 \pm 0.02 $& $ 0.78 \pm 0.08 $\\
		$ \Omega_{bb}^* \to \Xi_{bb} K $ & $ 0.61 \pm 0.04 $& $ 0.43 \pm 0.07 $\\
		$ \Omega_{bc}^* \to \Xi_{bc} K $ & $ 0.78 \pm 0.03 $& $ 0.55 \pm 0.06 $\\
		\hline 
		\hline 
	\end{tabular}
\end{table}
}
{
\renewcommand{\arraystretch}{1.2}
\begin{table}
	[H]\centering
	\caption{The numerical values of the coupling constants of the light vector mesons with the doubly heavy baryons.}\label{tab:4}
	\begin{tabular}
		{|c|c|c|c||c|c|c|} 
		\multicolumn{1}{c}{} & \multicolumn{3}{c}{Case of the general current} & \multicolumn{3}{c}{Case of the Ioffe current}\\
		\hline 
		\hline 
		Channel & $ g_1 $ & $ g_2 $ & $ g_3 $ & $ g_1 $ & $ g_2 $ & $ g_3 $\\
		\hline 
		$ \Xi_{cc}^* \to \Xi_{cc} \rho $ & $ 1.27 \pm 0.22 $ & $ 0.10 \pm 0.02 $ & $ 8.72 \pm 1.56 $ & $ 0.59 \pm 0.10 $ & $ 0.01 \pm 0.00 $ & $ 3.71 \pm 0.64 $\\
		$ \Xi_{bb}^* \to \Xi_{bb} \rho $ & $ 0.78 \pm 0.18 $ & $ 0.02 \pm 0.00 $ & $ 15.51 \pm 3.68 $ & $ 0.33 \pm 0.08 $ & $ 0.00 \pm 0.00 $ & $ 6.47 \pm 1.52 $\\
		$ \Xi_{bc}^* \to \Xi_{bc} \rho $ & $ 1.08 \pm 0.16 $ & $ 0.05 \pm 0.00 $ & $ 14.60 \pm 2.24 $ & $ 0.47 \pm 0.07 $ & $ 0.00 \pm 0.00 $ & $ 6.12 \pm 0.92 $\\
		$ \Omega_{cc}^* \to \Xi_{cc} K^* $ & $ 1.30 \pm 0.22 $ & $ 0.33 \pm 0.06 $  & $ 8.37 \pm 1.59 $ & $ 0.59 \pm 0.10 $ & $ 0.23 \pm 0.02 $ & $ 3.10 \pm 0.57 $\\
		$ \Omega_{bb}^* \to \Xi_{bb} K^* $ & $ 0.85 \pm 0.19 $ & $ 0.08 \pm 0.01 $ & $ 16.50 \pm 3.90 $ & $ 0.36 \pm 0.08 $ & $ 0.05 \pm 0.00 $ & $ 6.64 \pm 1.55 $\\
		$ \Omega_{bc}^* \to \Xi_{bc} K^* $ & $ 1.12 \pm 0.17 $ & $ 0.14 \pm 0.01 $ & $ 14.61\pm 2.32 $ & $ 0.48 \pm 0.07 $ & $ 0.09 \pm 0.00 $ & $ 5.81 \pm 0.91 $\\
		\hline 
		\hline 
	\end{tabular}
\end{table}
}
In Table \ref{tab:1}, we observe that the coupling constant, $ g $, of doubly heavy spin-3/2--spin-1/2 baryons with light pseudoscalar mesons are in good agreement with the ones for the case of the Ioffe current. In contrast, in Table \ref{tab:2}, one can see that the coupling constants, $ g_i $, of the said baryons with light vector mesons differ drastically from the ones for the case of the Ioffe current.
\section{Conclusion}\label{sec:4}
The discovery of $ \Xi _{cc}^{++} $ by SELEX and LHCb Collaborations stimulated theoretical and experimental studies for the investigation of the properties of other doubly heavy baryons in a new manner. Experimentally, the main attempt is focused on the discovery of doubly heavy baryons predicted by the quark model. Theoretically, the main effort is made to find promising decay channels, which can be potentially discovered in experiments. In this sense, one of the most important issues of doubly heavy baryons is the determination of the strong decay couplings among them. In the present work, we study the strong coupling constants of spin-3/2 to spin-1/2 transitions with pseudoscalar ($ \pi $ and $ K $) and vector ($ \rho $ and $ K^* $) mesons. The obtained results on these strong coupling constants can carry useful information not only about the internal structure of doubly heavy baryons, but also about the nonperturbative interaction of these objects. The results on these coupling constants can play a useful role in deeper studies about the properties of doubly heavy baryons. 
\appendix
\section{Distribution amplitudes for light mesons}\label{app:A}
In this section, for completeness, we collect the matrix elements $ \langle \mathcal M(q,\varepsilon) \vert \bar q(x) \Gamma _i q(0) \vert 0 \rangle $ and $ \langle \mathcal M(q,\varepsilon) \vert \bar q(x) \Gamma _i G_{\mu\nu} q(0) \vert 0 \rangle $ and the relevant distribution amplitudes for light mesons together with the most recent values for the DA parameters involved \cite{Ball1998, Ball1999, Ball1996, Ball2006, Ball19992, Ball2005}.
\prg
\textit{Pseudoscalar mesons}. Up to twist-4 accuracy, the matrix elements $ \langle \mathcal P(q,\varepsilon) \vert \bar q(x) \Gamma q(0) \vert 0 \rangle $ and $ \langle \mathcal P(q,\varepsilon) \vert \bar q(x) \Gamma G_{\mu\nu} q(0) \vert 0 \rangle $ are given as follows:
\begin{align}
	\bracket{\mathcal P (p)}{\bar q(x) \gamma _\mu \gamma _5 q (0)}{0} &= -i f _{\mathcal P} p _\mu \int _0^1 \dif u \ \e{i\bar u px} \bb{\varphi _{\mathcal P} (u) + \frac{1}{16} m _{\mathcal P}^2 x^2 \hat A (u)} 
	\nneq - \frac i 2 f _{\mathcal P} m _{\mathcal P}^2 \frac{x _\mu}{px} \int _0^1 \dif u \ \e{i\bar u px} \hat B(u) 
\end{align}
\begin{align}
	\bracket{\mathcal P (p)}{\bar q (x) i \gamma _5 q (0)}{0} &= \mu _{\mathcal P} \int _0^1 \dif u \ \e{i\bar u px} \varphi _P(u) 
\end{align}
\begin{align}
	\bracket{\mathcal P (p)}{\bar q (x) \sigma _{\alpha\beta} \gamma _5 q (0)}{0} &= \frac i 6 \mu _{\mathcal P} (1-\tilde \mu _{\mathcal P}^2) (p _\alpha x _\beta - p _\beta x _\alpha) \int _0^1 \dif u \ \e{i\bar u px} \varphi _\sigma (u)
\end{align}
\begin{align}
	\bracket{\mathcal P (p)}{\bar q (x) \sigma _{\mu\nu} \gamma _5 g _s G _{\alpha\beta} (vx) q (0)}{0} &= i \mu _{\mathcal P} \Big\{
	p _\alpha p _\mu \bb{g _{\nu\beta} - \frac{1}{px} (p _\nu x _\beta + p _\beta x _\nu)} 
	\nneq - p_\alpha p _\nu \bb{g _{\mu\beta} - \frac{1}{px} (p _\mu x _\beta + p _\beta x _\mu)} 
	\nneq - p _\beta p _\mu \bb{g _{\nu\alpha} - \frac{1}{px} (p _\nu x _\alpha + p _\alpha x _\nu)}
	\nneq + p _\beta p _\nu \bb{g _{\mu\alpha} - \frac{1}{px} (p _\mu x _\alpha + p _\alpha x _\mu)}
	\Big\}\nneq \times \int \mathcal D \alpha\ \e{i(\alpha _{\bar q} + v \alpha _g)px} \mathcal T (\alpha _i)
\end{align}
\begin{align}
	\bracket{\mathcal P (p)}{\bar q (x) \gamma _\mu \gamma _5 g _s G _{\alpha\beta} (vx) q(0)}{0} &= p _\mu (p _\alpha x _\beta - p _\beta x _\alpha) \frac{1}{px} f _{\mathcal P} m _{\mathcal P}^2 \int \mathcal D \alpha\ \e{i(\alpha _{\bar q} + v \alpha _g)px} \mathcal A _\parallel (\alpha _i) \nneq 
	+\Big\{
	p _\beta \bb{g _{\mu\alpha} - \frac{1}{px} (p _\mu x _\alpha 
		+ p _\alpha x _\mu)}
	\nneq - p _\alpha \bb{g _{\mu\beta} - \frac{1}{px} (p _\mu x _\beta + p _\beta x _\mu)}
	\Big\}  
	\nnhp{= +} \times  f _{\mathcal P} m _{\mathcal P} ^2 \int \mathcal D \alpha\ \e{i(\alpha _{\bar q} + v \alpha _g) px} \mathcal A _\perp (\alpha _i)
\end{align}
\begin{align}
	\bracket{\mathcal P (p)}{\bar q (x) \gamma _\mu i g _s G _{\alpha\beta} (vx) q(0)}{0} &= p _\mu (p _\alpha x _\beta - p _\beta x _\alpha) \frac{1}{px} f _{\mathcal P} m _{\mathcal P}^2 \int \mathcal D \alpha\ \e{i(\alpha _{\bar q} + v\alpha _g)px} \mathcal V _\parallel (\alpha _i)
	\nneq + \Big\{
	p _\beta \bb{g _{\mu\alpha} - \frac{1}{px} (p _\mu x _\alpha + p_\alpha x _\mu) } 
	\nneq - p _\alpha \bb{g _{\mu\beta} - \frac{1}{px} (p _\mu x _\beta + p _\beta x _\mu)}
	\Big\} 
	\nnhp{= + }\times f _{\mathcal P} m _{\mathcal P} ^2 \int \mathcal D \alpha \ \e{i(\alpha _{\bar q} + v\alpha _g)px} \mathcal V _\perp (\alpha _i) 
\end{align}
where
\begin{align}
	\mu _{\mathcal P} = f _{\mathcal P} \frac{m _{\mathcal P}^2}{m _{q _1} + m _{q _2}},\quad \tilde \mu _{\mathcal P} = \frac{m _{q _1} + m _{q _2}}{m _{\mathcal P}}
\end{align}
where $ m _{q _1} = m _u $ and $ m _{q _2} = m _d $ for the pion, and $ m _{q _1} = m _u $ and $ m _{q _2} = m _s $ for the kaon.
Here, $ \varphi _{\mathcal P} (u) $, $ \hat A (u) $, $ \hat B (u) $, $ \varphi _P (u) $, $ \varphi _\sigma (u) $, $ \mathcal T (\alpha _i) $, $ \mathcal A _\perp (\alpha _i) $, $ \mathcal A _\parallel (\alpha _i) $, $ \mathcal V _\perp (\alpha _i) $, and $ \mathcal V _\parallel (\alpha _i) $ are the distribution amplitudes of the pseudoscalar meson with definite twist. The relevant DAs are as follows:
\begin{align}
	\phi _{\mathcal P} (u) &= 6u\bar u \bb{
		1 + a_1^{\mathcal P} C_1(2u-1) + a_2^{\mathcal P} C_2^{3/2}(2u-1)	
	}\\
	\phi _P (u) &= 1 + \pp{
		30 \eta_3 - \frac 52 \frac 1{\mu _{\mathcal P}^2} 
	} C_2^{1/2} (2u-1) + \pp{
		-3\eta _3 w_3 -\frac{27}{20} \frac 1{\mu_{\mathcal P}^2} - \frac{81}{10} \frac1{\mu _{\mathcal P}^2} a_2^{\mathcal P}
	} C_4^{1/2}(2u-1)\\
	\phi _\sigma (u) &= 6u\bar u\bb{
		1 + \pp{
			5\eta_3 - \frac12 \eta_3 w_3 - \frac7{20} \mu_{\mathcal P}^2 - \frac35 \mu_{\mathcal P}^2 a_2^{\mathcal P}
		} C_2^{3/2}(2u-1)
	}\\
	\mathcal T(\alpha _i) &= 360 \eta_3 a_1 a_2 a_3^2 \bb{
		1 + w_3 \frac12 (7a_3-3)
	}
\end{align}
where the $ C_n^k(x) $ are the Gegenbauer polynomials and the values of the parameters inside the DAs at the renormalization scale of $ \mu = 1 {\rm\ GeV} $ are $ a_1^\pi = 0 $, $ a_2^\pi = 0.44 $, $ a_1^K = 0.06 $, $ a_2^K = 0.25 $, $ \eta_3 = 0.015 $, and $ w_3=-3 $ for the pion and $ w_3=-1.2 $ for the kaon. 
\prg
\textit{Vector mesons.} Up to twist-4 accuracy, the matrix elements $ \langle V(q,\varepsilon) \vert \bar q(x) \Gamma q(0) \vert 0 \rangle $ and $ \langle V(q,\varepsilon) \vert \bar q(x) \Gamma G_{\mu\nu} q(0) \vert 0 \rangle $ are given as follows:
\ba
\langle V(q,\varepsilon) \vert \bar q_1 (x) \gamma_\mu q_2(0) \vert 0 \rangle &=& f_V m_V (
\frac{\varepsilon^\lambda \cdot x}{q\cdot x} q_\mu \int_0^1 du\ e^{i\bar u q\cdot x} (\phi_2^\parallel (u) + \frac{m_V^2x^2}{16} \phi_4^\parallel(u)) 
\ar (\varepsilon_\mu^\lambda - q_\mu \frac{\varepsilon^\lambda \cdot x}{q\cdot x}) \int_0^1 du\ e^{i\bar u q\cdot x} \phi_3^\perp (u)
\ek \frac12 x_\mu \frac{\varepsilon^\lambda \cdot x}{(q\cdot x)^2} m_V^2 \int _0^1du\ e^{i\bar uq\cdot x} (\psi _4^\parallel(u) + \phi_2^\parallel (u) - 2 \phi_3^\perp (u))  
)
\ea
\ba
\langle V(q,\varepsilon) \vert \bar q_1(x) \gamma_\mu \gamma_5 q_2(0) \vert 0 \rangle = -\frac14 \epsilon _{\mu}^{\nu\alpha\beta} \varepsilon^\lambda_\nu q_\alpha x_\beta f_V m_V \int_0^1 du\ \psi_3^\perp (u)
\ea
\ba
\langle V(q,\varepsilon) \vert \bar q_1(x) \sigma_{\mu\nu} q_2 (0) \vert 0 \rangle &=& -if_V^T (
(\varepsilon^\lambda_\mu q_\nu - \varepsilon^\lambda_\nu q_\mu) \int_0^1du\ e^{i\bar u q \cdot x} \phi_2^\perp(u) + \frac{m_V^2x^2}{16} \phi_4^\perp(u) 
\ar \frac{\varepsilon \cdot x}{(q\cdot x)^2} (q_\mu x_\nu - q_\nu x_\mu) \int_0^1du \ e^{i\bar u q \cdot x} (\phi_3^\parallel (u) - \frac 12 \phi_2^\perp (u) - \frac 12 \psi_4^\perp(u))
\ar \frac 12 (\varepsilon^\lambda_\mu x_\nu - \varepsilon^\lambda_\nu x_\mu) \frac{m_V^2}{q\cdot x} \int_0^1 du\ e^{i\bar u q\cdot x} (\psi_4^\perp(u) - \phi_2^\perp(u))
)
\ea
\ba
\langle V(q,\varepsilon) \vert \bar q_1(x) \sigma_{\alpha\beta} g_s G_{\mu\nu} (ux) q_2(0) \vert 0 \rangle &=& f_V^T m_V^2 \frac{\varepsilon^\lambda \cdot x}{2q\cdot x} (q_\alpha q_\mu g_{\beta\nu}^\perp - q_\beta q_\mu g_{\alpha\nu}^\perp - q_\alpha q_\nu g_{\beta\mu}^\perp + q_\beta q_\nu g_{\alpha\mu}^\perp) 
\carp \int \mathcal D \alpha_i \ e^{i(\alpha_1 + u\alpha_3)q\cdot x} \mathcal T(\alpha _i) 
\ar f_V^T m_V^2 (q_\alpha \varepsilon_\mu^\lambda g_{\beta\nu}^\perp - q_\alpha \varepsilon_\nu^\lambda g_{\beta\mu}^\perp + q_\beta \varepsilon^\lambda_nu g_{\alpha\mu}^\perp) 
\carp \int \mathcal D\alpha_i \ e^{i(\alpha_1 + u \alpha_3)q\cdot x} \mathcal T_1^{(4)}(\alpha _i)
\ar f_V^T m_V^2 (q_\mu \varepsilon_\alpha^\lambda g_{\beta\nu}^\perp - q_\mu \varepsilon_\beta^\lambda g_{\alpha\nu}^\perp - q_\nu \varepsilon_\alpha^\lambda g_{\beta\mu}^\perp + q_\nu \varepsilon_\beta^\lambda g_{\alpha\mu}^\perp) 
\carp \int \mathcal D \alpha_i \ e^{i(\alpha_1 + u \alpha_3)q\cdot x} \mathcal T_2^{(4)}(\alpha _i)
\ar \frac{f_V^T m_V^2}{q\cdot x} (q_\alpha q_\mu \varepsilon^\lambda _\beta x_\nu - q_\beta q_\mu \varepsilon_\alpha^\lambda x_\nu - q_\alpha q_\nu \varepsilon^\lambda_\beta x_\mu + q_\beta q_\nu \varepsilon_\alpha^\lambda x_\mu) 
\carp \int \mathcal D \alpha_i \ e^{i(\alpha_1 + u \alpha_3)q\cdot x} \mathcal T_3^{(4)}(\alpha _i)
\ar \frac{f_V^T m_V^2}{q\cdot x} (q_\alpha q_\mu \varepsilon_\nu^\lambda x_\beta - q_\beta q_\mu \varepsilon_\nu^\lambda x_\alpha - q_\alpha q_\nu \varepsilon^\lambda_\mu x_\beta + q_\beta q_\nu \varepsilon^\lambda_\mu x_\alpha) 
\carp \int \mathcal D \alpha_i \ e^{i(\alpha_1 + u \alpha_3)q\cdot x} \mathcal T_4^{(4)}(\alpha _i)
\ea
\ba
\langle V(q,\varepsilon) \vert \bar q_1(x) g_s G_{\mu\nu} (ux) q_2(0) \vert 0 \rangle = -i f_V^T m_V (\varepsilon^\lambda_\mu q_\nu - \varepsilon^\lambda_\nu q_\mu) \int \mathcal D \alpha _i \ e^{i(\alpha_1 + u \alpha_3)q\cdot x} \mathcal S(\alpha _i)
\ea
\ba
\langle V(q,\lambda) \vert \bar q_1(x) g_s \tilde G_{\mu\nu} (ux)\gamma_5 q_2(0) \vert 0 \rangle = -i f_V^T m_V (\varepsilon_\mu^\lambda q_\nu - \varepsilon^\lambda_\nu q_\mu) \int\mathcal D \alpha _i \ e^{i(\alpha_1 + u\alpha_3) q\cdot x} \tilde{\mathcal S} (\alpha _i) 
\ea
\ba
\langle V(q,\varepsilon) \vert \bar q_1(x) g_s \tilde G_{\mu\nu} (ux) \gamma_\alpha \gamma_5 q_2(0) \vert 0 \rangle = f_V m_V q_\alpha (\varepsilon^\lambda_\mu q_\nu - \varepsilon^\lambda_nu q_\mu) \int\mathcal D\alpha_i\ e^{i(\alpha_1 + u \alpha_3)q\cdot x} \mathcal A(\alpha_i)
\ea
\ba
\langle V(q,\varepsilon) \vert \bar q_1(x) g_s G_{\mu\nu} (ux) i\gamma_\alpha q_2(0) \vert 0 \rangle = f_V m_V q_\alpha (\varepsilon^\lambda_\mu q_\nu - \varepsilon^\lambda_\nu q_\mu) \int \mathcal D \alpha _i \ e^{i(\alpha_1 + u \alpha_3)q\cdot x} \mathcal V(\alpha_i)
\ea
where $ \tilde G _{\mu\nu} = \frac12 \epsilon_{\mu\nu\alpha\beta} G^{\alpha\beta} $ is the dual gluon field strength tensor and $ \int \mathcal D\alpha_i = \int d\alpha_1 \ d\alpha_2 d\alpha_3 \delta(1-\alpha_1 - \alpha_2 - \alpha_3) $. Now we list the DAs.
\prg 
\textit{2-particle twist-2 DAs:}
\ba
\phiiiparu =
6 \baru (1 + \aipar C_{1}^{3/2}(\xi) + \aiipar C_{2}^{3/2}(\xi)) u
\ea
\ba
\phiiiperu = 
6 \baru (1 + \aiper C_{1}^{3/2}(\xi) + \aiiper C_{2}^{3/2}(\xi)) u
\ea
\textit{2-particle twist-3 DAs:}
\ba
\phiiiiparu &=& 
3 \xi^2 
+ (3 \aiper \xi (-1 + 3 \xi^2))/2 
+ ((15 \kappaiiiper)/2 - (3 \lambdaiiiper)/4) \xi (-3 + 5 \xi^2) 
\ar (3 \aiiper \xi^2 (-3 + 5 \xi^2))/2 
+ (5 \omegaiiiper (3 - 30 \xi^2 + 35 \xi^4))/8 
- (3 \fV (\mqi - \mqii) \xi (2 + 9 \aipar \xi \carp 2 \aiipar (11 - 30 \baru u) + (1 + 6 \aiipar + 3 \aipar) \ln(\baru) + (1 + 6 \aiipar - 3 \aipar) \ln(u)))/(2 \fVT \mV) 
\ar (3 \fV (\mqi + \mqii) (1 + 8 \aipar \xi + 3 \aiipar (7 - 30 \baru u) + (1 + 6 \aiipar + 3 \aipar) \xi \ln(\baru) - (1 + 6 \aiipar \ek 3 \aipar) \xi \ln(u)))/(2 \fVT \mV)
\ea
\ba
\psiiiiparu &=&
6 \baru (1 + C_{1}^{3/2}(\xi) (\aiper/3 + (5 \kappaiiiper)/3) - (C_{3}^{3/2}(\xi) \lambdaiiiper)/20 + C_{2}^{3/2}(\xi) (\aiiper/6 + (5 \omegaiiiper)/18)) 
\carp u - (3 \fV (\mqi - \mqii) (\baru (9 \aipar + 10 \aiipar \xi) u + (1 + 6 \aiipar + 3 \aipar) \baru \ln(\baru) - (1 + 6 \aiipar - 3 \aipar) 
\carp u \ln(u)))/(\fVT \mV) + (3 \fV (\mqi + \mqii) (\baru u (1 + 2 \aipar \xi + 3 \aiipar (7 - 5 \baru u)) + (1 + 6 \aiipar 
\ar 3 \aipar) \baru \ln(\baru) + (1 + 6 \aiipar - 3 \aipar) u \ln(u)))/(\fVT \mV)
\ea
\ba
\psiiiiperu &=&
6 \baru (1 + C_{1}^{3/2}(\xi) (\aipar/3 + (20 \kappaiiipar)/9) + C_{3}^{3/2}(\xi) (-\lambdaiiipar/8 + \lambdaiiipartilde/4) + C_{2}^{3/2}(\xi) (\aiipar/6 
\ar (5 \omegaiiipar)/12 - (5 \omegaiiipartilde)/24 + (10 \zetaiiipar)/9)) u - (6 \fVT (\mqi - \mqii) (\baru (9 \aiper + 10 \aiiper \xi) u + (1 
\ar 6 \aiiper + 3 \aiper) \baru \ln(\baru) - (1 + 6 \aiiper - 3 \aiper) u \ln(u)))/(\fV \mV) + (6 \fVT (\mqi + \mqii) (\baru u 
\carp (2 + 3 \aiper \xi + 2 \aiiper (11 - 10 \baru u)) + (1 + 6 \aiiper + 3 \aiper) \baru \ln(\baru) + (1 + 6 \aiiper - 3 \aiper) 
\carp u \ln(u)))/(\fV \mV)
\ea
\ba
\phiiiiperu &=&
(3 \aipar \xi^3)/2 + (3 (1 + \xi^2))/4 + (5 \kappaiiipar - (15 \lambdaiiipar)/16 + (15 \lambdaiiipartilde)/8) \xi (-3 + 5 \xi^2) 
\ar ((9 \aiipar)/112 + (15 \omegaiiipar)/32 - (15 \omegaiiipartilde)/64) (3 - 30 \xi^2 + 35 \xi^4) + (-1 + 3 \xi^2) ((3 \aiipar)/7 
\ar 5 \zetaiiipar) - (3 \fVT (\mqi - \mqii) (2 \xi + 2 \aiiper \xi (11 - 20 \baru u) + 9 \aiper (1 - 2 \baru u) + (1 + 6 \aiiper 
\ar 3 \aiper) \ln(\baru) - (1 + 6 \aiiper - 3 \aiper) \ln(u)))/(2 \fV \mV) + (3 \fVT (\mqi + \mqii) (2 + 9 \aiper \xi 
\ar 2 \aiiper (11 - 30 \baru u) + (1 + 6 \aiiper + 3 \aiper) \ln(\baru) 
\ar (1 + 6 \aiiper - 3 \aiper) \ln(u)))/(2 \fV \mV)
\ea
\textit{2-particle twist-4 DAs:}
\ba
\psiivparu &=&
1 + C_{3}^{1/2}(\xi) ((-9 \aipar)/5 - (20 \kappaiiipar)/3 - (16 \kappaivpar)/3) + C_{1}^{1/2}(\xi) ((9 \aipar)/5 + 12 \kappaivpar) 
\ar C_{3}^{1/2}(\xi) (-5 \thetaiipar + 10 \thetaipar) + (6 \fVT (\mqi - \mqii) (\xi + (\aiper (-1 + 3 \xi^2))/2 
\ar (5 \kappaiiiper (-1 + 3 \xi^2))/2 + (\aiiper \xi (-3 + 5 \xi^2))/2 + (5 \omegaiiiper \xi (-3 + 5 \xi^2))/6 
\ek (\lambdaiiiper (3 - 30 \xi^2 + 35 \xi^4))/16))/(\fV \mV) + C_{4}^{1/2}(\xi) ((-27 \aiipar)/28 - (15 \omegaiiipar)/8 
\ek (15 \omegaiiipartilde)/16 + (5 \zetaiiipar)/4) + C_{2}^{1/2}(\xi) (-1 - (2 \aiipar)/7 + (40 \zetaiiipar)/3) 
\ek (20 C_{2}^{1/2}(\xi) \zetaivpar)/3
\ea
\ba
\phiivparu &=&
(6 \baru \fVT (\mqi - \mqii) (-(C_{1}^{3/2}(\xi) ((82 \aiper)/5 + 10 \kappaiiiper)) + C_{3}^{3/2}(\xi) ((2 \aiper)/5 + (7 \lambdaiiiper)/54) 
\ar (2 C_{5}^{3/2}(\xi) \lambdaiiiper)/135 + C_{4}^{3/2}(\xi) (-2/315 + \aiiper/5 - \omegaiiiper/21) + 20 C_{2}^{3/2}(\xi) (10/189 
\ar \aiiper/3 - \omegaiiiper/21)) u)/(\fV \mV) + (6 \baru \fVT (\mqi + \mqii) (2 (3 + 16 \aiiper) + (10 C_{1}^{3/2}(\xi) (-\aiper 
\ar \kappaiiiper))/3 - (C_{3}^{3/2}(\xi) \lambdaiiiper)/10 + C_{2}^{3/2}(\xi) (-\aiiper + (5 \omegaiiiper)/9)) u)/(\fV \mV) 
\ar 30 \baru^2 (C_{1}^{5/2}(\xi) ((17 \aipar)/50 - \lambdaiiipar/5 + (2 \lambdaiiipartilde)/5) + (C_{2}^{5/2}(\xi) ((9 \aiipar)/7 + (7 \omegaiiipar)/6 
\ek (3 \omegaiiipartilde)/4 + \zetaiiipar/9))/10 + (4 (1 + \aiipar/21 + (10 \zetaiiipar)/9))/5) u^2 + 30 \baru^2 (C_{1}^{5/2}(\xi) ((2 \thetaiipar)/3 
\ek (8 \thetaipar)/15) + (20 \zetaivpar)/9) u^2 + (\fVT (\mqi - \mqii) ((-23 - 108 \aiiper - 54 \aiper + 5 u^2) \ln(\baru) 
\ek (-23 - 108 \aiiper + 54 \aiper + 5 \baru^2) \ln(u)))/(\fV \mV) + (24 \fVT (\mqi + \mqii) ((1 + 6 \aiiper 
\ar 3 \aiper) \baru^2 \ln(\baru) + (1 + 6 \aiiper - 3 \aiper) u^2 \ln(u)))/(\fV \mV) + 4 (\aipar - (40 \kappaiiipar)/3) ((11 
\ek 3 \xi^2)/8 - (2 - \baru) \baru^3 \ln(\baru) + (2 - u) u^3 \ln(u)) + 80 \psiiipar ((11 
- 3 \xi^2)/8 
\ek (2 - \baru) \baru^3 \ln(\baru) + (2 - u) u^3 \ln(u)) - 80 \omegaivpartilde ((\baru (21 - 13 \xi^2) u)/8 + \baru^3 (10 - 15 \baru 
\ar 6 \baru^2) \ln(\baru) + u^3 (10 - 15 u + 6 u^2) \ln(u)) + 2 (-2 \aiipar + 3 \omegaiiipar - (14 \zetaiiipar)/3) ((\baru (21 
\ek 13 \xi^2) u)/8 + \baru^3 (10 - 15 \baru + 6 \baru^2) \ln(\baru) + u^3 (10 - 15 u + 6 u^2) \ln(u))
\ea
\ba
\psiivperu &=&
1 + C_{1}^{1/2}(\xi) ((-3 \aiper)/5 + 12 \kappaivper) + (C_{5}^{1/2}(\xi) \lambdaiiiper)/3 + C_{4}^{1/2}(\xi) ((-3 \aiiper)/7 
\ek (5 \omegaiiiper)/4) + C_{3}^{1/2}(\xi) ((3 \aiper)/5 - 5 \kappaiiiper - 12 \kappaivper - \lambdaiiiper/3 + 5 ((-\thetaiiper - \thetaiipertilde)/2 + \thetaiper 
\ar \thetaipertilde)) + (\fV (\mqi + \mqii) (3 (1 + 6 \aiipar) + 3 \aipar C_{1}^{1/2}(\xi) + 5 C_{3}^{1/2}(\xi) (4 \kappaiiipar - (3 \lambdaiiipar)/4 
\ar (3 \lambdaiiipartilde)/2) + (15 C_{4}^{1/2}(\xi) (2 \omegaiiipar - \omegaiiipartilde))/4 + 5 C_{2}^{1/2}(\xi) (-3 \aiipar + 4 \zetaiiipar)))/(\fVT \mV) 
\ar C_{2}^{1/2}(\xi) (-1 + (3 \aiiper)/7 - 10 (\zetaivper + \zetaivpertilde)) - (6 \baru \fV (\mqi - \mqii) (9 \aipar + 10 \aiipar \xi)
\carp u)/(\fVT \mV) + (6 \fV (\mqi - \mqii) (-((1 + 6 \aiipar + 3 \aipar) \baru \ln(\baru)) + (1 + 6 \aiipar - 3 \aipar) 
\carp u \ln(u)))/(\fVT \mV) + (6 \fV (\mqi + \mqii) ((1 + 6 \aiipar + 3 \aipar) \baru \ln(\baru) + (1 + 6 \aiipar - 3 \aipar)
\carp u \ln(u)))/(\fVT \mV)
\ea
\ba
\phiivperu &=&
30 \baru^2 (2/5 + (4 \aiiper)/35 - (4 C_{3}^{5/2}(\xi) \lambdaiiiper)/1575 + C_{2}^{5/2}(\xi) ((3 \aiiper)/35 + \omegaiiiper/60) 
\ar C_{1}^{5/2}(\xi) ((3 \aiper)/25 + \kappaiiiper/3 - \lambdaiiiper/45 + (7 \thetaiiper)/30 - (3 \thetaiipertilde)/20 - \thetaiper/15 + \thetaipertilde/5) 
\ar (4 \zetaivper)/3 - (8 \zetaivpertilde)/3) u^2 + (-\aiper + 5 \kappaiiiper - 20 \phiiipertilde) ((\baru \xi (-11 + 3 \xi^2) u)/2 + 4 (2 - \baru) 
\carp \baru^3 \ln(\baru) - 4 (2 - u) u^3 \ln(u)) + ((-36 \aiiper)/11 - (252 \avqi)/55 - (140 \avqiii)/11 
\ar 2 \omegaiiiper) (-(\baru (-21 + 13 \xi^2) u)/8 + \baru^3 (10 - 15 \baru + 6 \baru^2) \ln(\baru) + u^3 (10 - 15 u + 6 u^2)
\carp \ln(u))
\ea
\textit{3-particle twist-3 DAs:}
\ba
\SS &=&
30 \alphaiii^2 (((-3 (\alphai^2 + (1 - \alphai - \alphaiii)^2))/2 + (1 - \alphaiii) \alphaiii) \psiiiper + (-6 \alphai (1 - \alphai - \alphaiii) 
\ar (1 - \alphaiii) \alphaiii) \psiiper + (1 - \alphaiii) \psioper - (-1 + 2 \alphai + \alphaiii) (((-3 + 5 \alphaiii) \thetaiiper)/2 + \alphaiii \thetaiper 
\ar \thetaoper))
\ea
\ba
\SSTilde &=&
30 \alphaiii^2 (((-3 (\alphai^2 + (1 - \alphai - \alphaiii)^2))/2 + (1 - \alphaiii) \alphaiii) \psiiipertilde + (-6 \alphai (1 - \alphai - \alphaiii) 
\ar (1 - \alphaiii) \alphaiii) \psiipertilde + (1 - \alphaiii) \psiopertilde - (\alphai - \alphaiii) (((-3 + 5 \alphaiii) \thetaiipertilde)/2 + \alphaiii \thetaipertilde 
\ar \thetaopertilde))
\ea
\ba
\VV =
360 \alphai (1 - \alphai - \alphaiii) \alphaiii^2 (\kappaiiipar + ((-3 + 7 \alphaiii) \lambdaiiipar)/2 + (-1 + 2 \alphai + \alphaiii) \omegaiiipar)
\ea
\ba
\AA =
360 \alphai (1 - \alphai - \alphaiii) \alphaiii^2 ((-1 + 2 \alphai + \alphaiii) \lambdaiiipartilde + ((-3 + 7 \alphaiii) \omegaiiipartilde)/2 + \zetaiiipar)
\ea
\ba
\TT =
360 \alphai (1 - \alphai - \alphaiii) \alphaiii^2 (\kappaiiiper + ((-3 + 7 \alphaiii) \lambdaiiiper)/2 + (-1 + 2 \alphai + \alphaiii) \omegaiiiper)
\ea
\textit{3-particle twist-4 DAs:}
\ba
\TTi =
120 \alphai (1 - \alphai - \alphaiii) \alphaiii ((-1 + 2 \alphai + \alphaiii) \phiiper + \phioper + (-1 + 3 \alphaiii) \phiiiper)
\ea
\ba
\TTii &=&
-30 \alphaiii^2 (-((-1 + 2 \alphai + \alphaiii) (((-3 + 5 \alphaiii) \psiiipertilde)/2 + \alphaiii \psiipertilde + \psiopertilde)) + ((-3 (\alphai^2 
\ar (1 - \alphai - \alphaiii)^2))/2 + (1 - \alphaiii) \alphaiii) \thetaiipertilde + (-6 \alphai (1 - \alphai - \alphaiii) + (1 - \alphaiii) \alphaiii) \thetaipertilde 
\ar (1 - \alphaiii) \thetaopertilde)
\ea
\ba
\TTiii =
-120 \alphai (1 - \alphai - \alphaiii) \alphaiii ((-1 + 3 \alphaiii) \phiiipertilde + (-1 + 2 \alphai + \alphaiii) \phiipertilde + \phiopertilde)
\ea
\ba
\TTiv &=&
30 \alphaiii^2 (-((-1 + 2 \alphai + \alphaiii) (((-3 + 5 \alphaiii) \psiiiper)/2 + \alphaiii \psiiper + \psioper)) + ((-3 (\alphai^2 
\ar (1 - \alphai - \alphaiii)^2))/2 + (1 - \alphaiii) \alphaiii) \thetaiiper + (-6 \alphai (1 - \alphai - \alphaiii) + (1 - \alphaiii) \alphaiii) \thetaiper 
\ar (1 - \alphaiii) \thetaoper)
\ea
where we have replaced $ \alpha_2 = 1-\alpha_1-\alpha_3 $ before the integration and we take $ \xi = \bar u $ since the second quark is at the point $ x = 0 $. The $ q _1 $ and $ q_2 $ indicate the quark components of the vector meson. The $ \rho $ meson has both light quarks, hence $ \mqi = \mqii = 0 $. The $ K^* $ meson has one strange quark and one light quark, thus $ \mqi = m_s $ but $ \mqii = 0 $. The derived DA parameters are given as follows:
\ba
\psioper = \zetaivper
\ea
\ba
\psiopertilde = \zetaivpertilde
\ea
\ba
\thetaoper = -(1/6) \kappaiiiper - (1/3) \kappaivper
\ea
\ba
\thetaopertilde = -(1/6) \kappaiiiper + (1/3) \kappaivper
\ea
\ba
\phioper = (1/6) \kappaiiiper + (1/3) \kappaivper
\ea
\ba
\phiopertilde = (1/6) \kappaiiiper - (1/3) \kappaivper
\ea
\ba
\phiiper = (9/44) \aiiper + (1/8) \omegaiiiper + (63/220) \avqi - (119/44) \avqiii
\ea
\ba
\phiipertilde = -(9/44) \aiiper + (1/8) \omegaiiiper - (63/220) \avqi - (35/44) \avqiii
\ea
\ba
\psiiper = (3/44) \aiiper + (1/12) \omegaiiiper + (49/110) \avqi - (7/22) \avqiii + (7/3) \avqv
\ea
\ba
\psiipertilde = -(3/44) \aiiper + (1/12) \omegaiiiper - (49/110) \avqi + (7/22) \avqiii + (7/3) \avqv
\ea
\ba
\psiiiper = -(3/22) \aiiper - (1/12) \omegaiiiper + (28/55) \avqi + (7/11) \avqiii + (14/3) \avqv
\ea
\ba
\psiiipertilde = (3/22) \aiiper - (1/12) \omegaiiiper - (28/55) \avqi - (7/11) \avqiii + (14/3) \avqv
\ea
\ba
\thetaipar = -(7/10) \aipar \zetaivpar
\ea
\ba
\thetaiipar = (7/5) \aipar \zetaivpar
\ea
\ba
\psiiipar = -(7/20) \aipar \zetaivpar
\ea
\ba
\thetaiper = -(21/10) \zetaivper \aiper
\ea
\ba
\thetaipertilde = (21/10) \zetaivper \aiper
\ea
\ba
\thetaiiper = (21/5) \zetaivper \aiper
\ea
\ba
\thetaiipertilde = -(21/5) \zetaivper \aiper
\ea
\ba
\phiiipertilde = -(21/20) \zetaivper \aiper
\ea
\ba
\avqi = -(10/3) \zetaivper
\ea
\ba
\avqiii = -\zetaivper
\ea
\ba
\avqv = 0
\ea
The numerical values for the DA parameters are given in Table \ref{tab:a-1}.
\def\birim#1{{\rm #1}}
{\renewcommand\arraystretch{1.5}
	\begin{table}[H]\centering
		\caption{The numerical values for the parameters in the DAs for vector mesons $ \rho $ and $ K^* $. The renormalization scale is $ \mu = 1 \ \birim{GeV} $.}\label{tab:a-1}
		\begin{tabular}{|c|c|c|c|c|c|c|c|c|c|}
			\hline 
			\hline 
			& $ f_V $ [GeV] & $ f_V^T $ [GeV] & $ m_V $ [GeV] & $ a_1^\parallel $ & $ a_1^\perp $ & $ a_2^\parallel $ & $ a_2^\perp $ & $ \zeta_3^\parallel $ & $ \tilde \lambda _3 ^\parallel $ \\
			\hline 
			$ \rho $ & 0.216 & 0.165 & 0.770 & 0 & 0 & 0.15 & 0.14& 0.03& 0\\
			$ K^* $ & 0.220 & 0.185 & 0.892 & 0.03 & 0.04 & 0.11 &0.1& 0.023&0.035 \\
			\hline 
			\hline
		\end{tabular}\vspace{5mm}
		\begin{tabular}{|c|c|c|c|c|c|c|c|c|c|c|c|c|c|}
			\hline 
			\hline 
			& $ \tilde \omega_3 ^\parallel $ & $ \kappa_3^\parallel $ & $ \omega_3^\parallel $ & $ \lambda _3 ^\parallel $ & $ \kappa _3^\perp $ & $ \omega_3^\perp $ & $ \lambda_3^\perp $ & $ \zeta_4^\parallel $ & $ \tilde\omega_4^\parallel $ & $ \zeta_4^\perp $ & $ \tilde \zeta _4^\perp $ & $ \kappa _4 ^\parallel $ & $ \kappa _4 ^\perp $\\
			\hline 
			$ \rho $ & --0.09 & 0 & 0.15 & 0 & 0 & 0.55 & 0 & 0.07 & --0.03 & --0.08 & --0.08 & 0 & 0  \\
			$ K^* $ & --0.07 & 0 & 0.1 & --0.008 & 0.003 & 0.3 & --0.025 & 0.02 & --0.02 & --0-.05& --0.05 & --0.025 & 0.013 \\
			\hline 
			\hline
		\end{tabular}\vspace{5mm}
	\end{table}
}
The light quark masses are taken to be zero, namely $ m_u = m_d = 0 $, and the mass of the strange quark at $ \mu = 1 \ \birim{GeV} $ is taken to be $ m_s = 0.137 \ \birim{GeV} $.
\section{Details of calculations in the theoretical part}\label{app:B}
In this section, we present some steps of the calculations needed in the theoretical analysis. In what follows, $ f(u) $ and $ \mathcal F (\alpha _i) $ denote generic 2- and 3-particle DAs, respectively, and we let $ K _i := K _i(m_Q\sqrt{-x^2})/(\sqrt{-x^2})^i $ and $ K_j := K _j(m_{Q'}\sqrt{-x^2})/(\sqrt{-x^2})^j $.
\prg 
\textit{Terms proportional to $ \mathcal O (\langle G \rangle^0) $:} For the sake of simplicity, we suppress the integral measures, $ \int du \int d^4x \ e^{i(p+\baru q)x}$, on the left-hand side. 
\ba
K_i K_j f(u) &\to& \frac i4 \frac{16\pi^2}{(2m_Q)^i (2m_{Q'})^j} (M^2)^{i+j} f(u_0) e^{-m_V^2/2M^2} \int _{(m_Q+m_{Q'})^2} ^{s_0} ds\ e^{-s/M^2} 
\carp \int d\alpha\ \alpha^{i-1} (1-\alpha)^{j-1} \delta(s - \frac{m_Q^2}{\alpha} - \frac{m_{Q'}^2}{1-\alpha})
\ea
\ba
x_\mu K_i K_j f(u) &\to& \frac{-i(u_0^{-1}p_\mu + q_\mu)}{M^2} \frac i4 \frac{16\pi^2}{(2m_Q)^i (2m_{Q'})^j} (M^2)^{i+j} f(u_0) e^{-m_V^2/2M^2} \int _{(m_Q+m_{Q'})^2} ^{s_0} ds\ e^{-s/M^2} 
\carp \int d\alpha\ \alpha^{i-1} (1-\alpha)^{j-1} \delta(s - \frac{m_Q^2}{\alpha} - \frac{m_{Q'}^2}{1-\alpha})
\ea
\ba
x_\mu x_\nu K_i K_j f(u) &\to& \frac{-(u_0^{-1}p_\mu+q_\mu) (u_0^{-1}p_\nu+q_\nu) - u_0^{-1}M^2 g_{\mu\nu}}{M^4} \frac i4 \frac{16\pi^2}{(2m_Q)^i (2m_{Q'})^j} (M^2)^{i+j} f(u_0) e^{-m_V^2/2M^2} 
\carp \int _{(m_Q+m_{Q'})^2} ^{s_0} ds\ e^{-s/M^2} \int d\alpha\ \alpha^{i-1} (1-\alpha)^{j-1} \delta(s - \frac{m_Q^2}{\alpha} - \frac{m_{Q'}^2}{1-\alpha})
\ea
\ba
x^2 K_i K_j f(u) &\to& \frac i4 \frac{16\pi^2}{(2m _Q)^i (2m_{Q'})^j} M^2 e^{-m_V^2/2M^2} f(u_0) \int _{(m_Q + m_{Q'})^2} ^{s_0} ds\ e^{-s/M^2} 
\carp \int d\alpha \ \alpha^{i-1} (1-\alpha)^{j-1} \delta(s-\frac{m_Q^2}{\alpha} - \frac{m_{Q'}^2}{1-\alpha}) (M^2)^{-3+i+j} 
\carp \frac{-4(-1+\alpha) (M^2 (-1+i+j) \alpha + m_Q^2) + 4\alpha m_{Q'}^2}{\alpha(\alpha-1)}
\ea
\ba
x_\mu x^2 K_i K_j f(u) &\to& \frac{-i(u_0^{-1}p_\mu+q_\mu)}{M^2} \frac i4 \frac{16\pi^2}{(2m _Q)^i (2m_{Q'})^j} M^2 e^{-m_V^2/2M^2} f(u_0) \int _{(m_Q + m_{Q'})^2} ^{s_0} ds\ e^{-s/M^2} 
\carp \int d\alpha \ \alpha^{i-1} (1-\alpha)^{j-1} \delta(s-\frac{m_Q^2}{\alpha} - \frac{m_{Q'}^2}{1-\alpha}) (M^2)^{-3+i+j} 
\carp \frac{-4(-1+\alpha) (M^2 (-1+i+j) \alpha + m_Q^2) + 4\alpha m_{Q'}^2}{\alpha(\alpha-1)}
\ea
\ba
x_\mu x_\nu x^2 K_i K_j f(u) &\to& \frac{-(u_0^{-1}p_\mu+q_\mu) (u_0^{-1}p_\nu + q _\nu)-u_0^{-1}M^2 g_{\mu\nu}}{M^4} \frac i4 \frac{16\pi^2}{(2m _Q)^i (2m_{Q'})^j} M^2 e^{-m_V^2/2M^2} f(u_0) 
\carp \int _{(m_Q + m_{Q'})^2} ^{s_0} ds\ e^{-s/M^2} \int d\alpha \ \alpha^{i-1} (1-\alpha)^{j-1} \delta(s-\frac{m_Q^2}{\alpha} - \frac{m_{Q'}^2}{1-\alpha}) 
\carp (M^2)^{-3+i+j} \frac{-4(-1+\alpha) (M^2 (-1+i+j) \alpha + m_Q^2) + 4\alpha m_{Q'}^2}{\alpha(\alpha-1)}
\ea
\ba
x^4 K_i K_j f(u) &\to& \frac i4 \frac{16\pi^2}{(2m_Q)^i (2m_{Q'})^j} M^2 e^{-m_V^2/2M^2} f(u_0) \int _{(m_Q+m_{Q'})^2}^{s_0} ds\ e^{-s/M^2} 
\carp \int d\alpha\ \alpha^{i-1} (1-\alpha)^{j-1} \delta(s - \frac{m_Q^2}{\alpha} - \frac{m_{Q'}^2}{1-\alpha}) \frac{16(M^2)^{-5+i+j}}{\alpha^2(-1+\alpha)^2} \carp(
M^4 (-2+i+j) (-1+i+j) (-1+\alpha)^2 \alpha^2 
\ar (-1+\alpha)^2 m_Q^2 (2M^2(-2+i+j) \alpha + m_Q^2)
\ek 2(-1+\alpha)\alpha (M^2(-2+i+j)\alpha + m_Q^2) m_{Q'}^2 + \alpha^2 m_{Q'}^4
)
\ea
\ba
x_\mu x^4 K_i K_j f(u) &\to& \frac{-i(u_0^{-1}p_\mu+q_\mu)}{M^2} \frac i4 \frac{16\pi^2}{(2m_Q)^i (2m_{Q'})^j} M^2 e^{-m_V^2/2M^2} f(u_0) \int _{(m_Q+m_{Q'})^2}^{s_0} ds\ e^{-s/M^2} 
\carp \int d\alpha\ \alpha^{i-1} (1-\alpha)^{j-1} \delta(s - \frac{m_Q^2}{\alpha} - \frac{m_{Q'}^2}{1-\alpha}) \frac{16(M^2)^{-5+i+j}}{\alpha^2(-1+\alpha)^2} \carp(
M^4 (-2+i+j) (-1+i+j) (-1+\alpha)^2 \alpha^2 
\ar (-1+\alpha)^2 m_Q^2 (2M^2(-2+i+j) \alpha + m_Q^2)
\ek 2(-1+\alpha)\alpha (M^2(-2+i+j)\alpha + m_Q^2) m_{Q'}^2 + \alpha^2 m_{Q'}^4
)
\ea
\ba
x_\mu x_\nu x^4 K_i K_j f(u) &\to& \frac{-(u_0^{-1}p_\mu+q_\mu) (u_0^{-1}p_\nu + q _\nu)-u_0^{-1}M^2 g_{\mu\nu}}{M^4} \frac i4 \frac{16\pi^2}{(2m_Q)^i (2m_{Q'})^j} M^2 e^{-m_V^2/2M^2} f(u_0) 
\carp \int _{(m_Q+m_{Q'})^2}^{s_0} ds\ e^{-s/M^2} \int d\alpha\ \alpha^{i-1} (1-\alpha)^{j-1} \delta(s - \frac{m_Q^2}{\alpha} - \frac{m_{Q'}^2}{1-\alpha}) 
\carp \frac{16(M^2)^{-5+i+j}}{\alpha^2(-1+\alpha)^2} (
M^4 (-2+i+j) (-1+i+j) (-1+\alpha)^2 \alpha^2 
\ar (-1+\alpha)^2 m_Q^2 (2M^2(-2+i+j) \alpha + m_Q^2)
\ek 2(-1+\alpha)\alpha (M^2(-2+i+j)\alpha + m_Q^2) m_{Q'}^2 + \alpha^2 m_{Q'}^4
)
\ea
\textit{Terms proportional to $ \mathcal O (\langle G \rangle^1) $:} For the sake of simplicity, we suppress the integral measures, $ \int du \int d^4x \int \mathcal D\alpha_i \ e^{i(p+(\alpha_1 + u\alpha_3) q)x} $, on the left-hand side.
\ba
K_i K_j f(u) \mathcal F (\alpha_i) &\to& \frac i4 \frac{16\pi^2}{(2m_Q)^i (2m_{Q'})^j} (M^2)^{i+j} e^{-m_V^2/2M^2} \int _{(m_Q + m_{Q'})^2} ^{s_0} ds\ e^{-s/M^2} 
\carp\int d\alpha\ \alpha^{i-1} (1-\alpha)^{j-1} \delta(s - \frac{m_Q^2}{\alpha} - \frac{m_{Q'}^2}{1-\alpha}) 
\carp \int _0^{1/2} d\alpha_1 \int _{1/2-\alpha_1}^{1-\alpha_1} d\alpha_3 \ f(\frac{u_0-\alpha_1}{\alpha_3})\frac{\mathcal F(\alpha _i)}{\alpha_3}
\ea
\ba
x_\mu K_i K_j f(u) \mathcal F (\alpha_i) &\to& \frac{-i(u_0^{-1}p_\mu+q_\mu)}{M^2} \frac i4 \frac{16\pi^2}{(2m_Q)^i (2m_{Q'})^j} (M^2)^{i+j} e^{-m_V^2/2M^2} 
\carp \int _{(m_Q + m_{Q'})^2} ^{s_0} ds\ e^{-s/M^2} \int d\alpha\ \alpha^{i-1} (1-\alpha)^{j-1} \delta(s - \frac{m_Q^2}{\alpha} - \frac{m_{Q'}^2}{1-\alpha}) 
\carp \int _0^{1/2} d\alpha_1 \int _{1/2-\alpha_1}^{1-\alpha_1} d\alpha_3 \ f(\frac{u_0-\alpha_1}{\alpha_3})\frac{\mathcal F(\alpha _i)}{\alpha_3}
\ea
\ba
x_\mu x_\nu K_i K_j f(u) \mathcal F (\alpha_i) &\to& \frac{-(u_0^{-1}p_\mu+q_\mu) (u_0^{-1}p_\nu + q _\nu)-u_0^{-1}M^2 g_{\mu\nu}}{M^4} \frac i4 \frac{16\pi^2}{(2m_Q)^i (2m_{Q'})^j} (M^2)^{i+j} e^{-m_V^2/2M^2} 
\carp \int _{(m_Q + m_{Q'})^2} ^{s_0} ds\ e^{-s/M^2} \int d\alpha\ \alpha^{i-1} (1-\alpha)^{j-1} \delta(s - \frac{m_Q^2}{\alpha} - \frac{m_{Q'}^2}{1-\alpha}) 
\carp \int _0^{1/2} d\alpha_1 \int _{1/2-\alpha_1}^{1-\alpha_1} d\alpha_3 \ f(\frac{u_0-\alpha_1}{\alpha_3})\frac{\mathcal F(\alpha _i)}{\alpha_3}
\ea
\ba
x^2 K_i K_j f(u) \mathcal F (\alpha _i) &\to& \frac i4 \frac{16\pi^2}{(2m_Q)^i (2m_{Q'})^j} e^{-m_V^2/2M^2} M^2 \int _{(m_Q + m_{Q'})^2} ^{s_0} ds \ e^{-s/M^2} 
\carp \int d\alpha \ \alpha^{i-1} (1-\alpha)^{j-1} \delta (s - \frac{m_Q^2}{\alpha} - \frac{m_{Q'}^2}{1-\alpha}) (M^2)^{-3+i+j} 
\carp \frac{-4(-1+\alpha)(M^2 (-1+i+j)\alpha + m_Q^2) + 4\alpha m_{Q'}^2}{(-1+\alpha)\alpha} 
\carp \int _0^{1/2} d\alpha _1 \int _{1/2-\alpha_1}^{1-\alpha_1} d\alpha_3 \ f(\frac{u_0-\alpha_1}{\alpha_3}) \frac{\mathcal F(\alpha _i)}{\alpha_3}
\ea
\ba
x_\mu x^2 K_i K_j f(u) \mathcal F (\alpha _i) &\to& \frac{-i(u_0p_\mu+q_\mu)}{M^2} \frac i4 \frac{16\pi^2}{(2m_Q)^i (2m_{Q'})^j} e^{-m_V^2/2M^2} M^2 \int _{(m_Q + m_{Q'})^2} ^{s_0} ds \ e^{-s/M^2} 
\carp \int d\alpha \ \alpha^{i-1} (1-\alpha)^{j-1} \delta (s - \frac{m_Q^2}{\alpha} - \frac{m_{Q'}^2}{1-\alpha}) (M^2)^{-3+i+j} 
\carp\frac{-4(-1+\alpha)(M^2 (-1+i+j)\alpha + m_Q^2) + 4\alpha m_{Q'}^2}{(-1+\alpha)\alpha} 
\carp \int _0^{1/2} d\alpha _1 \int _{1/2-\alpha_1}^{1-\alpha_1} d\alpha_3 \ f(\frac{u_0-\alpha_1}{\alpha_3}) \frac{\mathcal F(\alpha _i)}{\alpha_3}
\ea
\ba
x_\mu x_\nu x^2 K_i K_j f(u) \mathcal F (\alpha _i) &\to& \frac{-(u_0^{-1}p_\mu+q_\mu) (u_0^{-1}p_\nu + q _\nu)-u_0^{-1}M^2 g_{\mu\nu}}{M^4} \frac i4 \frac{16\pi^2}{(2m_Q)^i (2m_{Q'})^j} e^{-m_V^2/2M^2} M^2 
\carp \int _{(m_Q + m_{Q'})^2} ^{s_0} ds \ e^{-s/M^2} \int d\alpha \ \alpha^{i-1} (1-\alpha)^{j-1} \delta (s - \frac{m_Q^2}{\alpha} - \frac{m_{Q'}^2}{1-\alpha}) 
\carp (M^2)^{-3+i+j} \frac{-4(-1+\alpha)(M^2 (-1+i+j)\alpha + m_Q^2) + 4\alpha m_{Q'}^2}{(-1+\alpha)\alpha} 
\carp \int _0^{1/2} d\alpha _1 \int _{1/2-\alpha_1}^{1-\alpha_1} d\alpha_3 \ f(\frac{u_0-\alpha_1}{\alpha_3}) \frac{\mathcal F(\alpha _i)}{\alpha_3}
\ea
\ba
x^4 K_i K_j f(u) \mathcal F(\alpha_i) &\to& \frac i4 \frac{16\pi^2}{(2m_Q)^i (2m_{Q'})^j} M^2 e^{-m_V^2/2M^2} \int _{(m_Q+m_{Q'})^2}^{s_0} ds\ e^{-s/M^2} 
\carp \int d\alpha\ \alpha^{i-1} (1-\alpha)^{j-1} \delta(s-\frac{m_Q^2}{\alpha} - \frac{m_{Q'}^2}{1-\alpha}) \frac{16(M^2)^{-5+i+j}}{(-1+\alpha)^2\alpha^2} \carp(
M^4(-2+i+j) (-1+i+j) (-1+\alpha)^2 \alpha^2 
\ar (-1+\alpha)^2 m_Q^2 (2M^2 (-2+i+j) \alpha + m_Q^2) 
\ek 2(-1+\alpha) \alpha (M^2 (-2+i+j) \alpha +m_Q^2) m_{Q'}^2 + \alpha m_{Q'}^4 
) 
\carp\int _0^{1/2} d\alpha _1 \int _{1/2-\alpha_1}^{1-\alpha_1} d\alpha_3\ f(\frac{u_0-\alpha_1}{\alpha_3}) \frac{\mathcal F(\alpha_i)}{\alpha_3}
\ea
\ba
x_\mu x^4 K_i K_j f(u) \mathcal F(\alpha_i) &\to& \frac{-i(u_0^{-1}p_\mu+q_\mu)}{M^2} \frac i4 \frac{16\pi^2}{(2m_Q)^i (2m_{Q'})^j} M^2 e^{-m_V^2/2M^2} \int _{(m_Q+m_{Q'})^2}^{s_0} ds\ e^{-s/M^2} 
\carp \int d\alpha\ \alpha^{i-1} (1-\alpha)^{j-1} \delta(s-\frac{m_Q^2}{\alpha} - \frac{m_{Q'}^2}{1-\alpha}) \frac{16(M^2)^{-5+i+j}}{(-1+\alpha)^2\alpha^2} \carp (
M^4(-2+i+j) (-1+i+j) (-1+\alpha)^2 \alpha^2 
\ar (-1+\alpha)^2 m_Q^2 (2M^2 (-2+i+j) \alpha + m_Q^2) 
\ek 2(-1+\alpha) \alpha (M^2 (-2+i+j) \alpha +m_Q^2) m_{Q'}^2 + \alpha m_{Q'}^4 
) 
\carp \int _0^{1/2} d\alpha _1 \int _{1/2-\alpha_1}^{1-\alpha_1} d\alpha_3\ f(\frac{u_0-\alpha_1}{\alpha_3}) \frac{\mathcal F(\alpha_i)}{\alpha_3}
\ea
\ba
x_\mu x_\nu x^4 K_i K_j f(u) \mathcal F(\alpha_i) &\to& \frac{-(u_0^{-1}p_\mu+q_\mu) (u_0^{-1}p_\nu + q _\nu)-u_0^{-1}M^2 g_{\mu\nu}}{M^4} \frac i4 \frac{16\pi^2}{(2m_Q)^i (2m_{Q'})^j} M^2 e^{-m_V^2/2M^2} 
\carp \int _{(m_Q+m_{Q'})^2}^{s_0} ds\ e^{-s/M^2} \int d\alpha\ \alpha^{i-1} (1-\alpha)^{j-1} \delta(s-\frac{m_Q^2}{\alpha} - \frac{m_{Q'}^2}{1-\alpha}) 
\carp \frac{16(M^2)^{-5+i+j}}{(-1+\alpha)^2\alpha^2} (
M^4(-2+i+j) (-1+i+j) (-1+\alpha)^2 \alpha^2 
\ar (-1+\alpha)^2 m_Q^2 (2M^2 (-2+i+j) \alpha + m_Q^2) 
\ek 2(-1+\alpha) \alpha (M^2 (-2+i+j) \alpha +m_Q^2) m_{Q'}^2 + \alpha m_{Q'}^4 
) 
\carp \int _0^{1/2} d\alpha _1 \int _{1/2-\alpha_1}^{1-\alpha_1} d\alpha_3\ f(\frac{u_0-\alpha_1}{\alpha_3}) \frac{\mathcal F(\alpha_i)}{\alpha_3}
\ea
where
\begin{align}
	u_0 := \frac{M_1^2}{M_1^2+M_2^2}
\end{align}
For terms containing $ q\cdot x $, we perform the following operation:
\ba 
(q\cdot x)^n f(u) \to \begin{cases}
	(-i \frac{\del}{\del u})^n f(u), & n > 0\\
	i^{-n} \int _0^u dv _n \cdots \int _0^{v_3} dv_2 \int _0^{v_2}dv_1 f(v_1), & n < 0
\end{cases}
\ea 
\begin{align}
	(q\cdot x)^n \mathcal F (\alpha _i) \to \begin{cases}
		(\frac iu \frac{\del}{\del \alpha _3})^n \mathcal F (\alpha _i), & n > 0\\
		(-iu)^{-n} \int _0^{\alpha _3} d\alpha _3^{(n)} \cdots \int _0^{\alpha_3^{(3)}}d\alpha_3^{(2)} \int _0^{\alpha _3 ^{(2)}}d\alpha_3^{(1)} \mathcal F (\alpha _1, 1-\alpha_1-\alpha_3^{(1)},\alpha_3^{(1)}), & n < 0
	\end{cases}
\end{align}
\bibliography{paper}
\end{document}